\documentclass[aps,pra,reprint,superscriptaddress,notitlepage]{revtex4-1}
\usepackage{amsmath}
\usepackage{amssymb}
\usepackage{amsthm}
\usepackage{bbm}
\usepackage{graphicx}
\usepackage{hyperref}
\usepackage{physics}
\usepackage{times}

\RequirePackage{color}

\newcommand{\e}[1]{\{#1\}}

\hypersetup{
  colorlinks=true,linkcolor=blue,citecolor=blue,
  filecolor=blue,urlcolor=blue,breaklinks=true
}

\begin{document}

\title{Narrow peaks of full transmission in simple quantum graphs}

\author{A. Drinko}
\email{alexandredrinko@gmail.com}
\affiliation{
  Programa de P\'os-Gradua\c{c}\~ao em Ci\^encias/F\'{i}sica,
  Universidade Estadual de Ponta Grossa,
  84030-900 Ponta Grossa, PR, Brazil
}

\author{F. M. Andrade}
\email{fmandrade@uepg.br}
\affiliation{
  Programa de P\'os-Gradua\c{c}\~ao em Ci\^encias/F\'{i}sica,
  Universidade Estadual de Ponta Grossa,
  84030-900 Ponta Grossa, PR, Brazil
}
\affiliation{
  Departamento de Matem\'{a}tica e Estat\'{i}stica,
  Universidade Estadual de Ponta Grossa,
  84030-900 Ponta Grossa, PR, Brazil
}

\author{D. Bazeia}
\email{bazeia@fisica.ufpb.br}
\affiliation{
  Departamento de F\'{i}sica,
  Universidade Federal da Para\'{i}ba
  58051-900 Jo\~{a}o Pessoa, PB, Brazil
}
\date{\today}

\begin{abstract}
This work deals with quantum graphs, focusing on the transmission properties they engender. We first select two simple diamond graphs, and two hexagonal graphs in which the vertices are all of degree 3, and investigate their transmission coefficients. In particular, we identified regions in which the transmission is fully suppressed. We also considered the transmission coefficients of some series and parallel arrangements of the two basic graphs, with the vertices still preserving the degree 3 condition, and then identified specific series and parallel compositions that allow for windows of no transmission. Inside some of these windows, we found very narrow peaks of full transmission, which are consequences of constructive quantum interference. Possibilities of practical use as the experimental construction of devices of current interest to control and manipulate quantum transmission are also discussed.
\end{abstract}

\maketitle

\section{Introduction}

In the past 20 years, quantum graphs
\cite{AoP.274.76.1999,AP.55.527.2006} have been used
to describe the behavior of quantum particles in idealized physical
networks.
The interest is due to the richness of the subject, which can be related
to a variety of issues in physical and mathematical sciences.
For instance, it has been simulated experimentally in  microwave
networks \cite{PRE.69.056205.2004} and it is also possible to synthesize
quantum nanowire networks \cite{NM.3.380.2004,NL.8.4523.2008}.
From the fundamental point of view, quantum graphs have became a test
bed for studying different aspects in quantum mechanics and, due to the
complex nature of the problem, the development of a unique method that
holds for all graphs is difficult.
Fortunately, however, there are some techniques developed in the
literature that are able to deal with this problem
\cite{Book.2012.Berkolaiko}.
Among the several methods to deal with quantum graphs, an interesting
one is the Green's function approach, first proposed in
\cite{JPA.36.545.2003} and further explored in
\cite{PR.647.1.2016,PRA.98.062107.2018}.
In this work we shall deal with specific scattering properties of
quantum graphs, which are identified by two leads and sets of vertices
and edges, to be described in the next section.
The focus is mainly on the transmission properties of simple graphs,
owing to the possibility of applications of physical interest.
We investigate the global transmission amplitude of
quantum graphs as a function of the wave number of the incident signal
using the Green's function approach developed in
\cite{PR.647.1.2016,PRA.98.062107.2018}.
In particular, closer attention is given to the search for a new effect,
which somehow remind us of the Braess paradox \cite{UOR.12.258.1968},
and to the
possibility to identify regions of wave numbers where the
transmission coefficient increases significantly. The study will lead to the
identification of peaks of quantum interference of very narrow
width, similar to Feshbach resonances \cite{AoP.5.357.1958}, in distinct
arrangements of the basic structures to be studied in this work.

As one knows, the Braess paradox was first discussed by Braess
in 1968 \cite{UOR.12.258.1968}, and further studied in
Refs. \cite{TS.39.446.2005,PRL.101.128701.2008}.
Originally, it showed that adding an extra road to a congested road
traffic network to improve traffic flow may sometimes have the reverse
effect, impeding the flow.
The paradox was
also discussed in several other contexts \cite{PRL.108.076802.2012,PRB.88.245417.2013,N.352.699.1991,
AJP.71.479.2003,PRE.90.042915.2014}, in particular
in \cite{PRL.108.076802.2012}, in which the quantum
transport in mesoscopic networks with two and three branches revealed the
transport inefficiency, confirmed by a scanning-probe
experiment using a biased tip that modulates the conductance variation
in terms of the tip voltage and position. It also appeared in \cite{PRB.88.245417.2013} in a
quantum ring of finite width, and in \cite{PRE.90.042915.2014} in the context
of two quantum dots that are coupled together.
Besides exploring global transmission properties of quantum graphs,
we also concentrate on the presence of very narrow peaks of full
transmission. The narrowness of the peaks reminds us of Feshbach resonances
\cite{AoP.5.357.1958}, which, in the context of quantum graphs, were
investigated before in Ref. \cite{APPA.124.1087.2013} in a ring graph
with edges with unequal sizes.
Motivated by the above reasonings, in this work we follow the
interpretation that the addition of a new path may under specific
conditions enlarge the complexity of the system, opening new
possibilities that may include unexpected responses.
The effects that we are interested here appear in the search for the
global transmission related to simple quantum graphs in the presence of
quantum interference.

The investigation deals with quantum graphs, but in this work we
consider simple graphs that are formed by arrangements of ideal leads,
edges and vertices.
This means that neither the vertices nor the leads and edges allow for
vanishing of the quantum probability and, in this sense, any linear
arrangement of leads, edges and vertices is trivial, since it gives full
global transmission (see below).
To go further on out of the trivial situation,
we  have to consider the possibility of a vertex being connected by a
triple junction, in the form of a Y-shaped configuration, which is
usually called a vertex of degree 3.
In this case, the signal reaching a vertex has the possibility to
reflect and return, and two possible paths of transmission, and this
introduces nontrivial quantum effects that can appear in the global
transmission coefficient of the graph.
Of course, there is a diversity of possibilities of constructing graphs
with vertices connected by two, three, four and more edges, and so in
this work we consider the two simple possibilities of diamond and
hexagonal arrangements.
In the hexagonal case, we pay special attention to two arrangements,
composed of two leads, six vertices and eight edges, with
all the vertices having degree 3.
We consider the hexagonal graphs in order to eliminate effects that could
appear due to the presence of vertices with higher degrees, which seem
to prefer backscattering \cite{JPA.40.14181.2007}.
We call this the degree 3 condition, and use it to simplify the current
investigation.
The presence of vertices of degree 3 leads us to two distinct graphs
that can be used to build arrangements that gives interesting responses,
which can be of practical use in the construction of simple devices with
important quantum properties.
The two graphs to be considered in this work are hexagonal graphs with
some internal structure, and this will also be of current interest,
since hexagons are important to tile the plane to lead to hexagonal
structures with vertices of degree 3, which are important part in
nanotubes \cite{Book.1998.Saito} and in graphene sheets
\cite{RMP.81.109.2009,NL.7.3394.2007}.

In order to deal with the above issues and implement the investigation,
we organize the work as follows.
In the next Sec. \ref{sec:graph}, the main properties of quantum graphs
are  described, and there one concentrates mainly on the global
transmission of simple graphs via the Green's function approach.
In Sec. \ref{sec:sg} we introduce some simple graphs and describe
and compare their global transmission coefficients.
This investigation allows that we identify another effect, which is
periodic and appears under the presence of quantum complexity.
Also, in Sec. \ref{sec:gc} we deal with some simple composition of two
distinct graphs and study some simple series and parallel arrangements
of them.
This will lead us to the presence of narrow peaks of full transmission,
so in Sec. \ref{sec:r} we further investigate the issue to search for
the presence of very narrow peaks of quantum interference.
In Sec. \ref{sec:end} we end the work, adding some comments and
conclusions, paying further attention to the possibility of using the
results of the work to applications of current interest to
quantum transport.

\section{Procedure}
\label{sec:graph}
In this section, we review some concepts of graphs as used in this
paper.
In particular, we deal with quantum effects and the use of the Green's
function approach for the calculation of the global transmission
properties of quantum graphs, paying closer attention to the case of
graphs with simple arrangements of leads, edges, and vertices.

\subsection{Quantum graphs}
A graph $G(V,E)$ consists of a set of vertices $V(G)=\{1,\ldots,n\}$
and a set of edges  $E(G)=\{e_{1},\ldots,e_{l}\}$
\cite{Book.2010.Diestel}.
The graph is described in terms of the adjacency matrix $A(G)$
of dimension $n \times n$ where the $ij$th element is defined by
\begin{equation}
  A_{ij}(G)=
  \begin{cases}
    1, & \text{if } \{i,j\}\in E(G),\\
    0, & \text{otherwise}.
  \end{cases}
\end{equation}
The degree of a vertex $i$ is defined as
$d_{i}=\sum_{j=1}^{n}A_{ij}(G)$.
We denote the set of neighbors of a vertex $i$ by
$E_{i}=\{j: e_{s}= \{i,j\}\in E(G)\}$
and the set of neighbors of $i$ but with
the vertices $\{k_1, \ldots , k_{d_{i}}\}$ excluded by
$E_{i}^{k_1,\ldots,k_{d_i}} = E_{i}\setminus \{k_1, \ldots , k_{d_{i}}\}$.
A metric graph $\Gamma(V,E)$ is a graph in which it is assigned a
positive length $\ell_{e_{s}}\in(0,+\infty)$ to each edge.
When a single ended edge $e_{s}$ is taken as semi-infinite
($\ell_{e_{s}} = + \infty$), it is called a {\it lead}.
A quantum graph is a metric graph in which is possible to define a
Schr\"odinger operator along with appropriated boundary conditions at
the vertices \cite{AP.55.527.2006}.
In general, the Schr\"odinger operator along the edge $\{i,j\}$ has the
form
\begin{equation}\label{potential}
H_{ij}=-\frac{\hbar^2}{2m}\frac{d^2}{dx^2}+V_{ij}(x),
\end{equation}
where $V_{ij}(x)$ is the corresponding potential. In this sense, we can
model distinct edges with the inclusion of different potentials;
in particular, one can add a square well which will modify the
transmission through the edge and so the global transmission through the
graph.
In this work, however, we shall take $V_{ij}(x)=0$, that is, we shall
use the free Schr\"odinger operator.

\subsection{The Green's function approach}
\label{sec:gfa}

In the context of quantum graphs, the exact scattering Green's function
for a quantum particle of fixed energy $E=\hbar^2k^2/2m$, with initial
position $x_i$ in the lead $e_i$ and final position $x_n$ in the lead
$e_{n}$, is given by a sum over all the scattering paths connecting
the points $x_i$ and $x_n$, where each path is weighted by the product
of the scattering amplitudes gained along the path \cite{PR.647.1.2016}.
The reflection and transmission amplitudes, $r_{i}$ and $t_{i}$, at the
vertex $i$, are determined through the boundary conditions defined at
the vertex $i$.
With the help of the adjacency matrix of the graph, it was shown in
\cite{PRA.98.062107.2018} that this sum over the paths can be written in
the form
\begin{equation}
  G_{\Gamma_{in}} = \frac{m}{i \hbar^2 k}
  T_{\Gamma_{in}}(k)  e^{i k (x_{i}+x_{n})},
\end{equation}
where
\begin{equation}
  \label{eq:Tglobal}
  T_{\Gamma_{in}}(k) = \sum_{j \in E_{i}} t_{i} A_{ij} p_{ij}^{(n)},
\end{equation}
is the \textit{transmission amplitude}.
The $p_{ij}^{(n)}$ is the family of paths between the vertices $i$ and
$j$, which are given by
\begin{equation}
  \label{eq:pij}
  p_{ij}^{(n)}
  = z_{ij} r_{j} p_{ji}^{(n)}
  +\sum_{l \in {E_{j}^{i,n}}} z_{ij}  t_{j} A_{jl} p_{jl}^{(n)}
  + z_{ij} t_{n} \delta_{jn},
\end{equation}
with $z_{ij}= e^{i k \ell_{\{i,j\}}}$.
The family $p_{ji}^{(n)}$ is given by the same expression above, but
with the swapping of indices $i$ and $j$.
Then, in each vertex $i$ we associated one $p_{ij}^{(n)}$ for every
$j \in E_{i}$.
In this work, we shall employ the above approach to determine the
\textit{transmission coefficient} $|T_{\Gamma_{in}}(k)|^{2}$ for
different quantum graphs and then discuss their properties.
A useful quantity in our discussion is the difference between the
transmission coefficients of two quantum graphs, $\Gamma$ and $\Gamma'$,
and will be represented by
  \begin{equation}
  \Delta_{\Gamma \Gamma'}(k) =
  \left| T_{\Gamma}(k) \right|^2 -
  \left| T_{\Gamma'}(k) \right|^2.
\end{equation}
Here, we shall consider equilateral quantum graphs where the length of
all the edges are the same and set them to $\ell$, such that
$z_{ij}=z=e^{i k \ell}$.

\subsection{The quantum amplitudes}
Let us now discuss the possible boundary conditions.
As one knows, a common vertex condition used is the so-called
$\delta$-type
condition defined by \cite{PRL.74.3503.1995}
\begin{align}
  \label{eq:bc_delta}
  \psi_{\e{i,j}} = {} & \varphi_{j},
  \qquad \forall i \in E_{j},
  \nonumber \\
  \sum_{i \in E_j} \psi_{\e{i,j}}'
  = {} & \alpha_{j}\;\varphi_{j},
\end{align}
where $\varphi_{j}$ is the value of the wave function at the vertex $j$
and $\alpha_{j}$ is a real parameter related to the strength of the
$\delta$-type interaction.
The prime in \eqref{eq:bc_delta} represents the derivative, which should
be taken in the outgoing direction, i.e., from the vertex into the edges
or leads.
Using this boundary condition the quantum amplitudes have the form \cite{PRA.98.062107.2018,PR.647.1.2016}
\begin{align}
  \label{eq:r_delta}
  r_{j}(k) = {} &
  \frac
  {
    \alpha_{j} -(d_{j}-2) i k}
  {i k d_{j} - \alpha_{j}},\\
  \label{eq:t_delta}
  t_{j}(k) = {} &
  \frac{2ik}
  {i k d_{j} - \alpha_{j}}.
\end{align}
Among the choices for the value of $\alpha_j$, an interesting one is
$\alpha_j=0, \forall j$.
In this case, we are considering no barrier at the vertices, resulting
in the so-called Neumann-Kirchhoff boundary condition.
As a result, the quantum amplitudes have the property of being
independent of $k$,
\begin{equation}
r_{j} =  \frac{2}{d_{j}}-1, \qquad
t_{j} =  \frac{2}{d_{j}},
\end{equation}
showing that the reflection amplitude increases with the increase of
the degree of the vertex.
When a Neumann-Kirchhoff boundary condition is used at a vertex of
degree 2, the transmission amplitude is equal to 1 and the vertex
becomes an ordinary point joining the edges.
Such vertices are called Neumann vertices \cite{AP.55.527.2006}.
In what follows, we shall adopt vertices of degree 3 with the
Neumann-Kirchhof boundary condition.
In this case the reflection and transmission amplitudes are explicitly
given by $r_{j}=-1/3$ and $t_{j}=2/3$, respectively.

With the above conditions, we shall then be considering quantum graphs
with ideal leads and edges, and with vertices that obey the
Neumann-Kirchhoff boundary conditions.
This is the simplest possibility, and we shall mainly consider vertices
of degree 3, to avoid accounting for effects due to vertices of
different degrees.

\section{Simple graphs}
\label{sec:sg}

\begin{figure}[!t]
  \centering
  \includegraphics*[width=0.9\columnwidth]{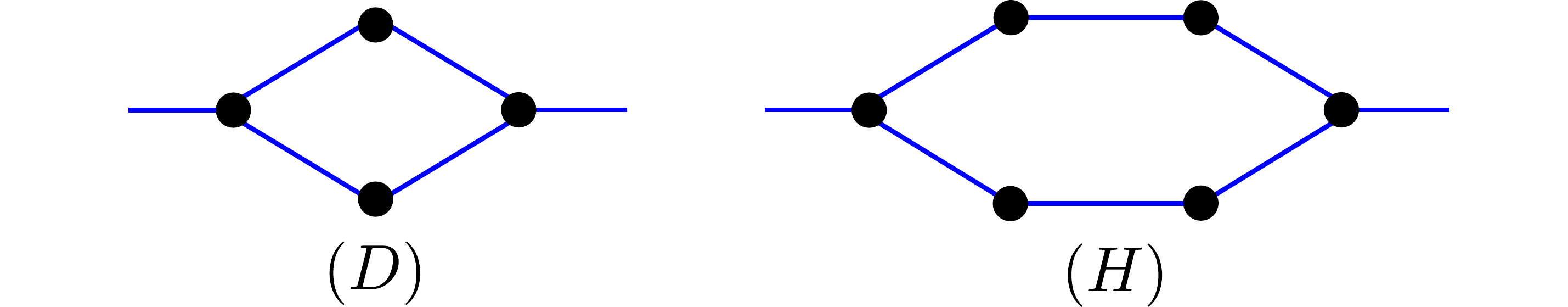}
  \caption{(Color online)
   Illustration of the diamond and hexagonal graphs.
  }
  \label{fig:fig1}
\end{figure}

\begin{figure}[!h]
  \centering
  \includegraphics*[width=\columnwidth]{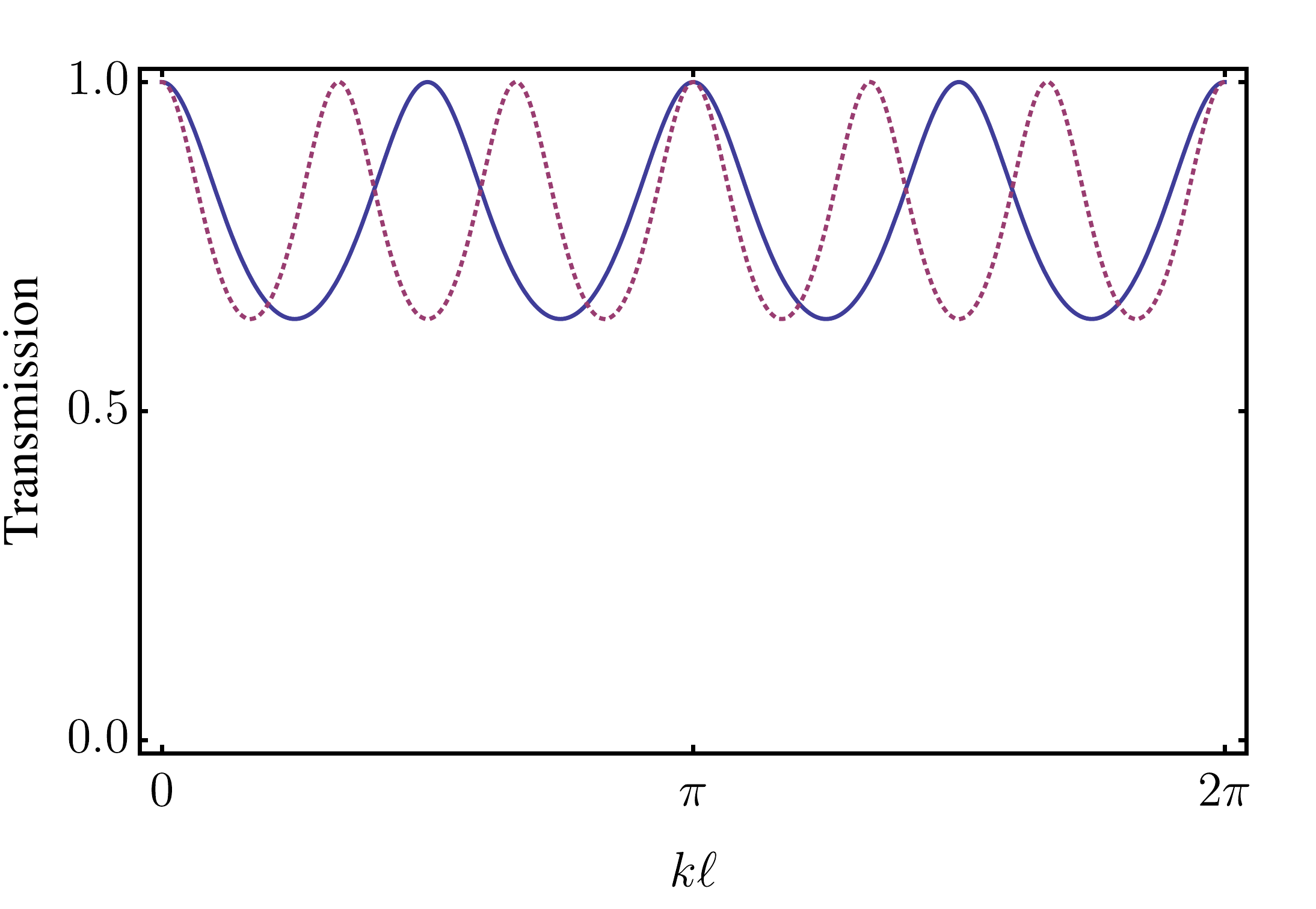}
  \caption{(Color online)
    Transmission coefficients of the diamond (blue solid line) and
    hexagonal (violet dotted line) graphs shown in Fig. \ref{fig:fig1}.}
  \label{fig:fig2}
\end{figure}

Let us now concentrate on simple graphs.
One uses the symbol $\bullet$ to represent vertices, and straight line
segments to stand for the edges and leads, and first considers the graph
${\color{blue}-}{\!\bullet\!}{\color{blue}-}$
with two leads -- one at the left and the other at the right of the
vertex.
As we have already commented, this is trivial and gives
$|T_{{\color{blue}-}{\!\bullet\!}{\color{blue}-}}(k)|^2=1$, since we are
considering a Neumann vertex.
If we go further and consider the graph
${\color{blue}-}{\!\bullet\!}{\color{blue}-}{\!\bullet\!}{\color{blue}-}$
we also get full transmission,
$|T_{{\color{blue}-}{\!\bullet\!}{\color{blue}-}{\!\bullet\!}{\color{blue}-}}(k)|^2=1$.
In view of this, we have to consider vertices with higher degrees
leading to more complex graphs to open the possibility of having
nontrivial transmission effects, so we depict the diamond ($D$) and the
hexagonal ($H$) graphs that are shown in Fig.~\ref{fig:fig1}.
In this case, the transmission amplitudes can be written as
\begin{align}
  T_{D}(k) =  {}
  & \frac{8 z^2}{9-z^4},\\
  T_{H}(k) = {} & \frac{8 z^{3}}{9-z^{6}},
\end{align}
and the corresponding transmissions coefficients are shown in Fig.~\ref{fig:fig2}.

The transmission coefficients are no longer  nontrivial, and the effects 
are due to the presence of the left and right vertices of degree 3.
Here we notice that since we are considering Neumann vertices, the
difference between the diamond and hexagonal graphs depicted in
Fig. \ref{fig:fig1} is only due to the difference between the two
internal paths, which is at the ratio $2/3$, as it nicely appears when
one accounts for the difference between the periodicity of the
transmission of the diamond and the hexagonal graphs that appear in
Fig. \ref{fig:fig2}.

\begin{figure}[t!]
  \centering
  \includegraphics*[width=0.9\columnwidth]{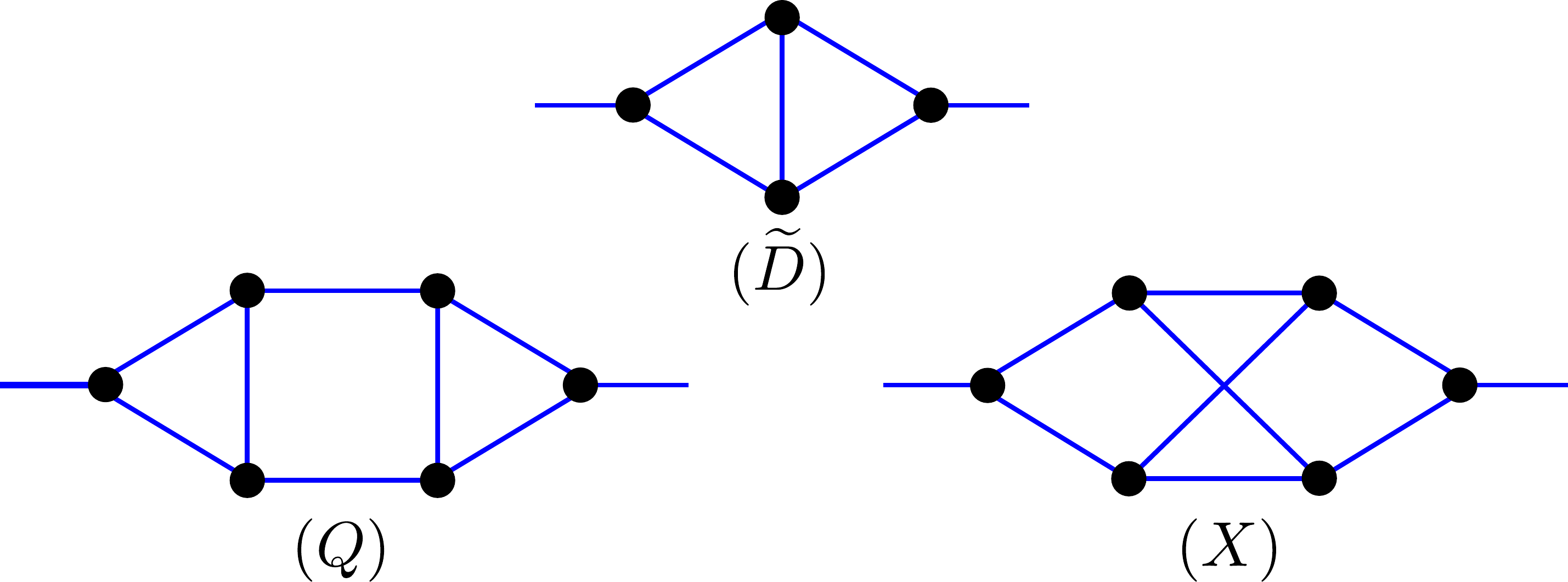}
  \caption{(Color online)
    Graphs with vertices of degree 3 in the diamond and hexagonal families.}
  \label{fig:fig3}
\end{figure}

\begin{figure}[t!]
  \centering
  \includegraphics*[width=\columnwidth]{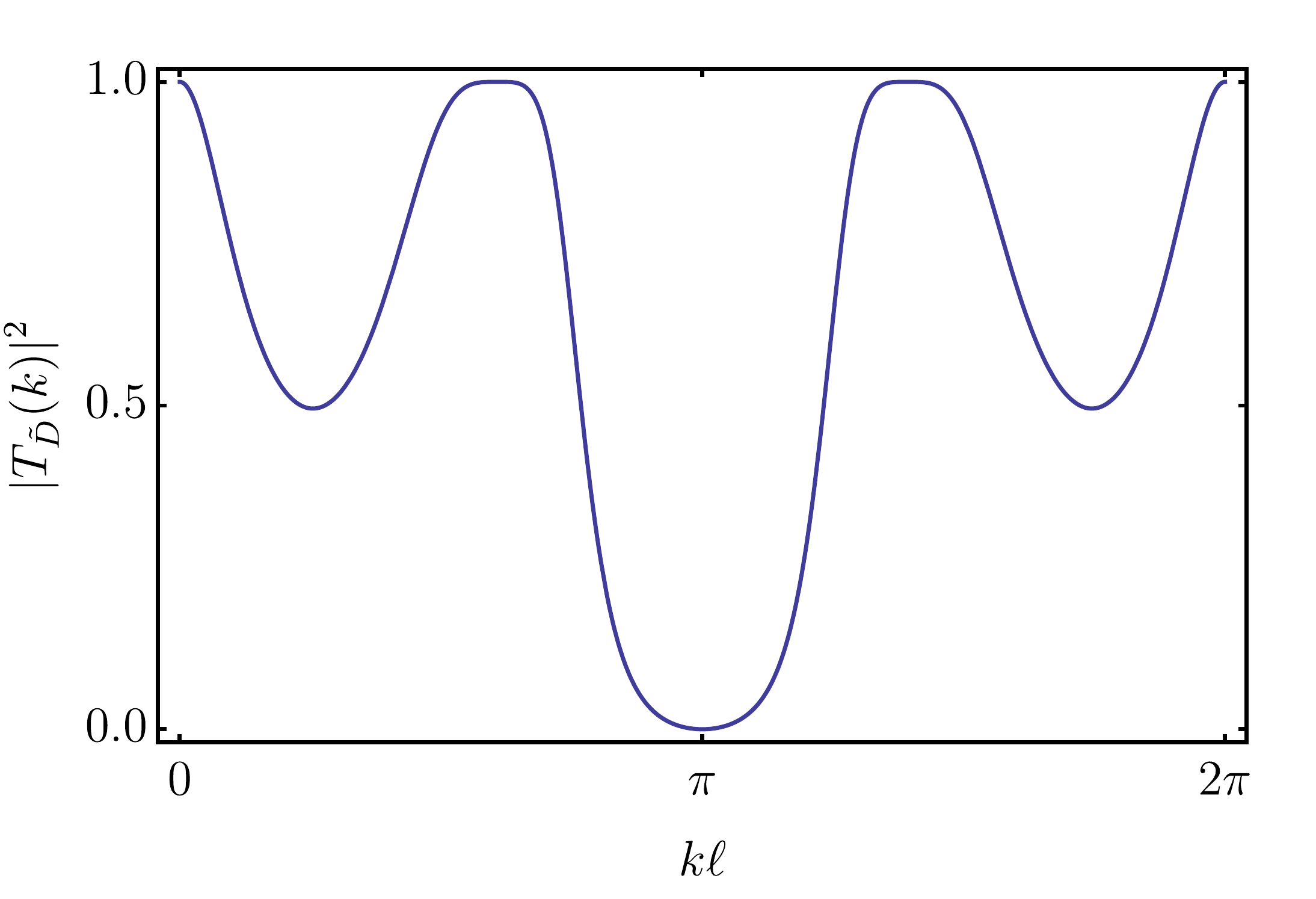}\!\!\!
  \includegraphics*[width=\columnwidth]{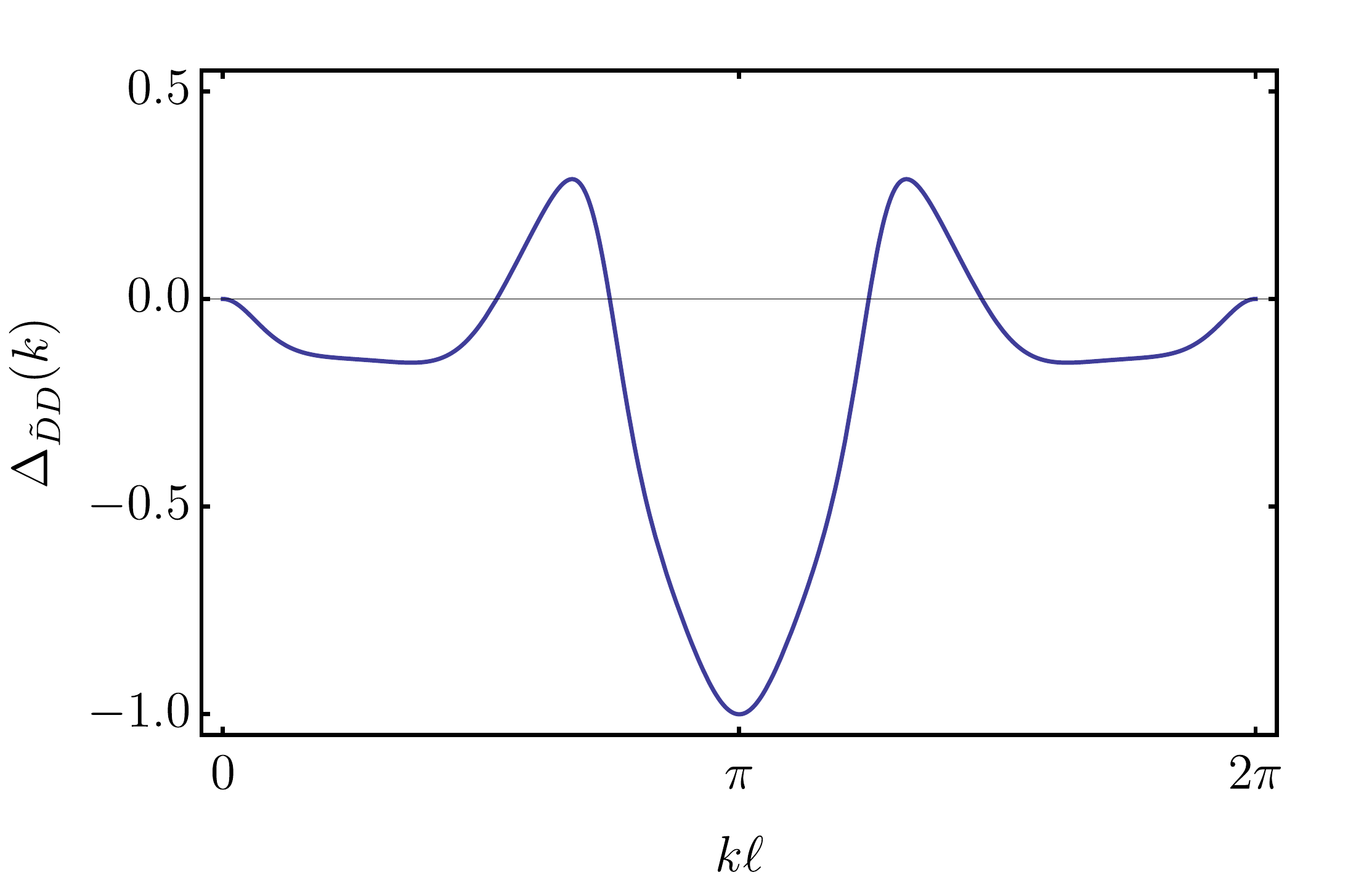}
  \caption{(Color online)
    Transmission coefficient of the ${\widetilde D}$ graph (top) and the difference between the transmission coefficients of the
    $\widetilde D$ and $D$ graphs.}
  \label{fig:fig4}
\end{figure}

\begin{figure}[t!]
  \centering
  \includegraphics*[width=\columnwidth]{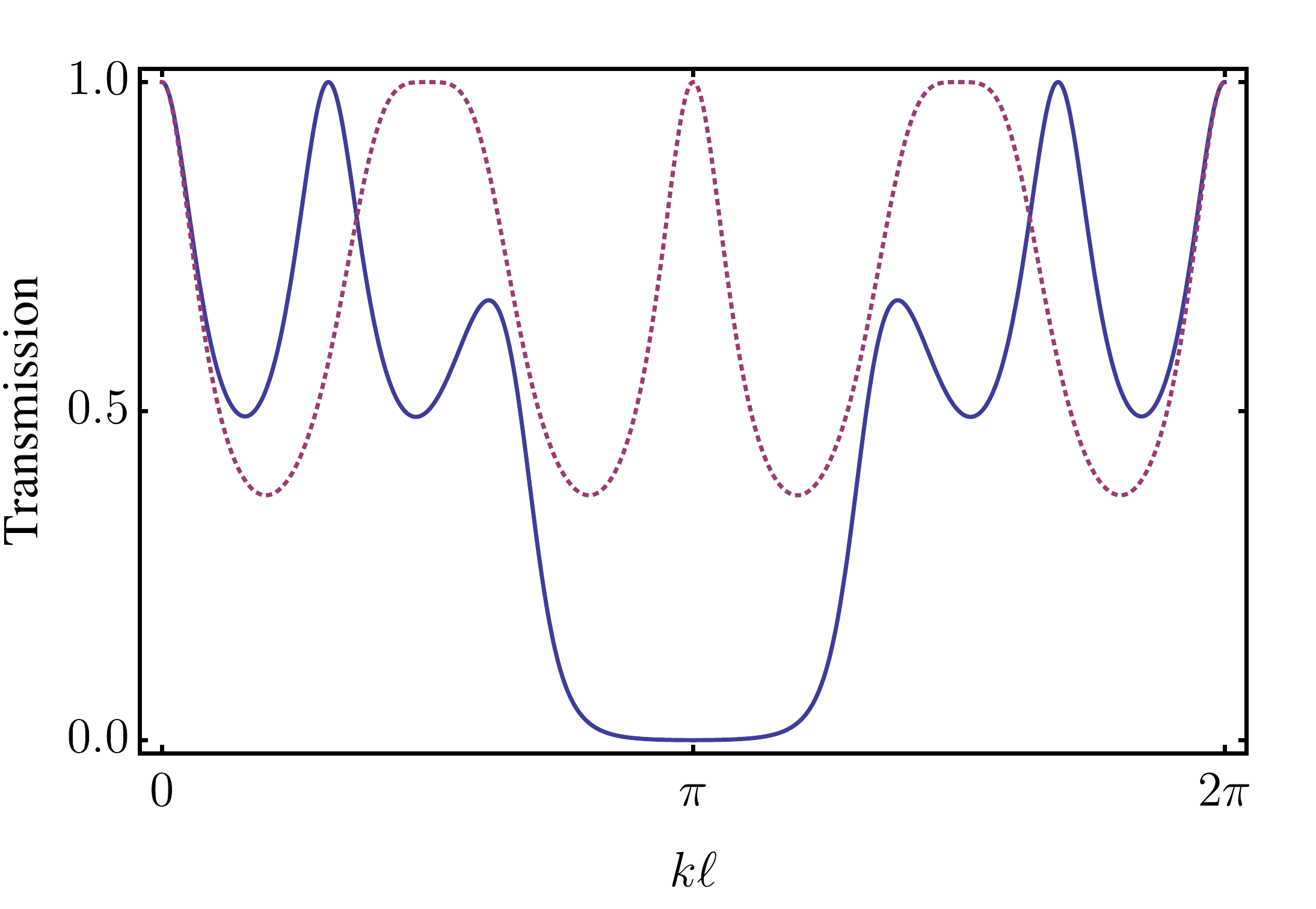}
  \caption{(Color online)
    Transmission coefficients of the two structures, the hexagonal
    square (blue solid line) and crossed (violet dotted line)
    arrangements $Q$ and $X$ that appear in Fig. \ref{fig:fig3}.
  }
  \label{fig:fig5}
\end{figure}

Since the presence of vertices of degree 3 induces nontrivial
transmission, we then focus on this and consider the new graphs which
are depicted in Fig. \ref{fig:fig3}.
They are all constructed with vertices of degree 3 in the diamond and
hexagonal families. Since there is only one graph in the diamond family,
we cannot compare the transmission behavior of two distinct diamond
graphs with vertices of degree 3.
However, if we relax the degree 3 condition and consider the two diamond
graphs that are depicted in Figs. \ref{fig:fig1} and \ref{fig:fig3}, we
see that the diamond graph ${\widetilde D}$ of  Fig. \ref{fig:fig3} has
an extra edge which is not present in the diamond graph $D$ of
Fig. \ref{fig:fig1}, so it seems that the transmission through it would
always be greater than the other one, which was already calculated
above. To see how this works, we calculate the transmission coefficient
of the
${\widetilde D}$ graph, which is given by
\begin{equation}
T_{\widetilde D}(k)=\frac{16 z^2(1+z)}{27+9z+6z^2-6z^3-z^4-3z^5}.
\end{equation}
We then depict the difference $\Delta_{\widetilde D D}(k)$ in Fig. \ref{fig:fig4} and the result shows that the transmission through
${\widetilde D}$ is not always greater than the one through $D$.
The effect shows that the addition of an extra edge in the graph $D$ to
make it the ${\widetilde D}$ graph, does not always improve the
transmission probability.
We believe that this is a manifestation of the quantum complexity that
appears in the ${\widetilde D}$ graph.
However, since the two graphs $D$ and ${\widetilde D}$ have different
numbers of edges, and vertices with different degrees, we cannot
separate how these effects contribute to the final transmissions.
Due to this, from now on we concentrate on the transmission coefficient
of the two graphs $Q$ and $X$ of the hexagonal family that are depicted
in Fig. \ref{fig:fig3} to compare their properties.
We stress that all the vertices of these two hexagonal graphs have
degree 3 and, also, they have the same number of edges and vertices.
This means that our results will be contaminated neither by effects of
vertices of different degrees nor by effects of different number of
vertices and edges.
The transmission amplitudes for these two graphs are given by
\begin{align}
  T_{Q} (k) = {}
  & \frac{32 z^{3}(1+z)}{(9+4z^2+3z^4)(9-3z+z^2-3z^3)},\label{eq:tS}\\
  T_{X} (k) = {}
  & \frac{64 z^3}{81+9z^2-17z^4-9z^6}\label{eq:tX},
\end{align}
and the corresponding transmission coefficients are displayed in Fig. \ref{fig:fig5},
unveiling interesting nontrivial properties which we discuss below.

As we noted, the two transmission coefficients are periodic so we
display them in Fig. \ref{fig:fig5} for the wave number in the interval
of periodicity of the $Q$ graph.
We also observe that the transmission coefficient of the $Q$ graph is
more complex  than the other one.
More importantly, it may vanish in a large interval which we call the
suppression band, inside its interval of periodicity.
The results also show that there are regions in $k$ space, where the
transmission is more or less significant for the $Q$ than the $X$
graphs.
That is, $|T_{Q}(k)|^2$ may be greater or smaller than
$|T_{X}(k)|^2$, depending on the interval in $k$ space in
consideration.

\begin{figure}[t!]
  \centering
  \includegraphics*[width=\columnwidth]{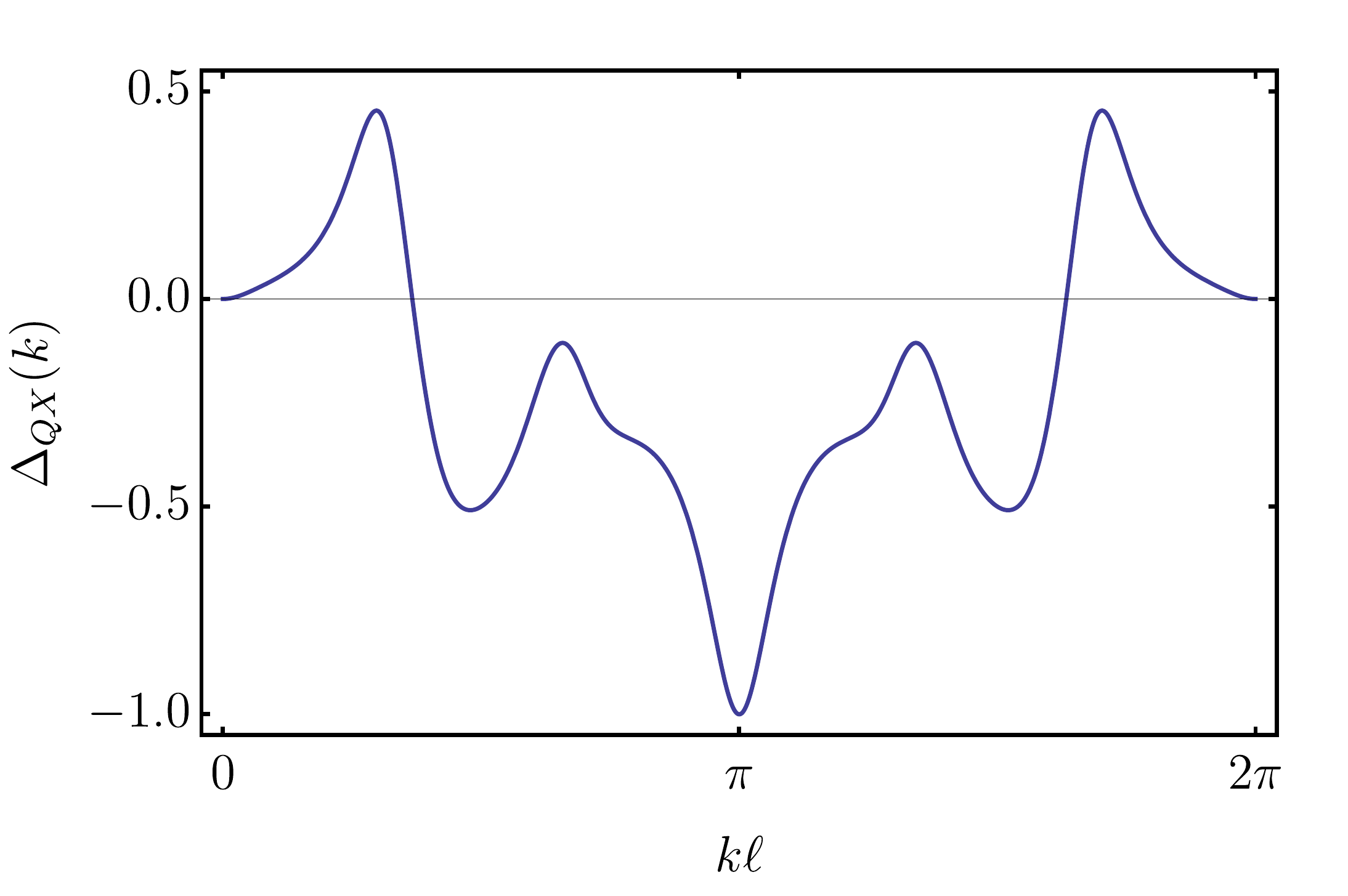}
  \caption[]{(Color online)
    Difference between the transmission coefficients of the $Q$ and
    $X$ graphs.}
  \label{fig:fig6}
\end{figure}

These results open the possibility to study the presence of an effect
that is similar to the one that appeared above, with the two diamond
graphs, but now with the vertices with the same degree.
Here we depict the difference $\Delta_{QX}(k)$
between the two coefficients, to see regions where one is higher than
the other. This is shown in Fig. \ref{fig:fig6}, and we can observe that
for $k\ell$ in the interval $1.15215 < k\ell < 5.13103 $ the
transmission of the $X$ graph is higher than the transmission of the $Q$
graph. However, for $0<k\ell<1.15215$ and for $5.13103<k\ell<2\pi$ one sees
that the transmission for the $Q$ graph is greater than the one for the
$X$ graph, and this is a manifestation of the quantum complexity
related to these two graphs.
In regard to the numerical results, all of them were obtained using the
commercial Mathematica software by using standard techniques with
precision of twelve decimal digits, and we decided to use only five
digits in the final results.

To get further insight on the result depicted in Fig. \ref{fig:fig6},
let us now briefly discuss the transmission of the signal at the
classical level: one supposes that a classical signal enters the graph
at the left (right) lead and leaves it at the right (left) lead; one
notices from Fig. \ref{fig:fig3} that for the $Q$ graph the shortest
trajectory requires three steps, and that there are two distinct
possibilities; for the $X$ graph, the shortest trajectory also requires
three steps, but now there are four distinct possibilities.
This suggests that the classical flux through the $X$ graph seems to be
more efficient than the other one.
However, the problem is more complex than this, because of the presence
of reflection at the vertices.
For instance, the next shortest trajectory for the $Q$ graph requires
four steps, and there are four possibilities; for the $X$ graph there is
no trajectory with four steps. In fact, the $X$ graph has only
trajectories with odd number of steps, whereas the $Q$ graph has both
even and odd number of steps.
Thus, even classically, it is hard to decide which of the two graphs is
more efficient to transmit the signal.
At the quantum level the problem is harder, because of the quantum
interference, and the result displayed in Fig. \ref{fig:fig6} shows that
the answer depends on the wave number.
Since the two graphs have the same number of edges and vertices and all
the vertices have degree $3$, it is the topological difference between
these two graphs that leads to different transmission coefficients,
which depends on the wave number and the quantum complexity involved in
the problem.

The result in Fig. \ref{fig:fig6} suggests that both the $Q$ and $X$
graphs are complex enough to give rise to other effects of current
interest.
They then motivate us to go further and explore other possibilities.
In the current investigation we explore the fact that the $Q$ graph
engenders a band of no transmission in $k$ space, as it is easily
identified from the blue solid line depicted in Fig. \ref{fig:fig5}.
This is the suppression band, and it is an interesting
quantum effect that can be used in different applications.
The simplest possibility is to use it to block the passage of
a signal through the quantum graph, which can be seen as a device of
direct interest to the construction of tools that allow for the control
and manipulation of quantum transmission probability.
Yet more interesting is to see the two quantum graphs as two independent
quantum devices, which can be used for the construction of others,
composed devices, and this will be investigated in the next section.

\begin{figure}[t!]
  \centering
    \includegraphics*[width=\columnwidth]{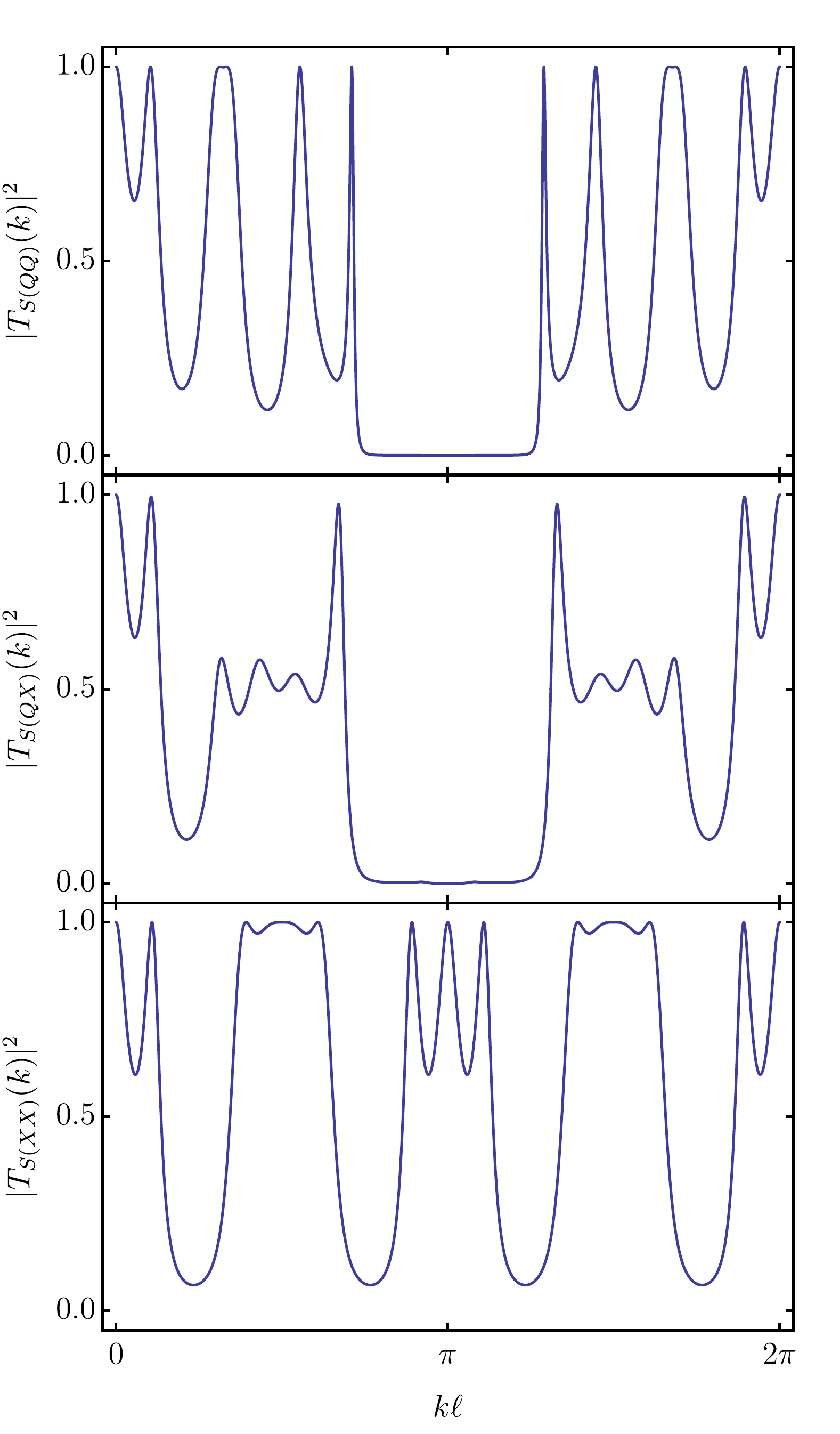}
  \caption[]{(Color online)
    Transmission coefficients of the three
    compound structures, the series arrangements $S(QQ)$, $S(QX)$, and $S(XX)$,
    depicted from top to bottom, respectively.
  }
  \label{fig:fig7}
\end{figure}

\section{Graph circuitry}
\label{sec:gc}

Let us now use the two quantum devices, the $Q$ and the $X$ graphs,
to build compound structures and study their transmission properties in
light of the above investigation. 

\subsection{Simple series circuits}

We first consider the series composition, of the forms $S(QQ)$,
$S(QX)$, $S(XQ)$, and $S(XX)$, where $S(QX)$ indicates the series
composition of the graph $Q$ with the graph $X$, keeping the degree 3
condition of the vertices, which in the series arrangement occurs very
naturally.
We then calculate and display the transmission coefficient for all the
cases in Fig. \ref{fig:fig7}.
Although the square and crossed graphs are different, the compound
transmission does not depend on the order one chooses each other, so we
say that $S(QX)=S(XQ)$.
We compare the transmission displayed with the blue solid line in
Fig. \ref{fig:fig5} with the one in the top panel in Fig. \ref{fig:fig7}
to see that the series composition of two square graphs enlarges a bit
the suppression band in $k$ space around $k\ell =\pi$, so it is a
bit more efficient to block the passage of a signal.
On the other hand, the violet dotted line that appears in
Fig. \ref{fig:fig5} and the bottom panel in Fig. \ref{fig:fig7} show the
appearance of extra maxima in the transmission coefficient of the
$S(XX)$ graph. 
This composition also deepens the main minima, approaching them to
suppression.
The composition $S(QX)$ which appears in the middle panel in
Fig. \ref{fig:fig7} is also interesting: it shows an almost invisible
substructure in the suppression band, and this suggests that we further
explore this effect.

\begin{figure}[t!]
  \centering
 \includegraphics*[width=\columnwidth]{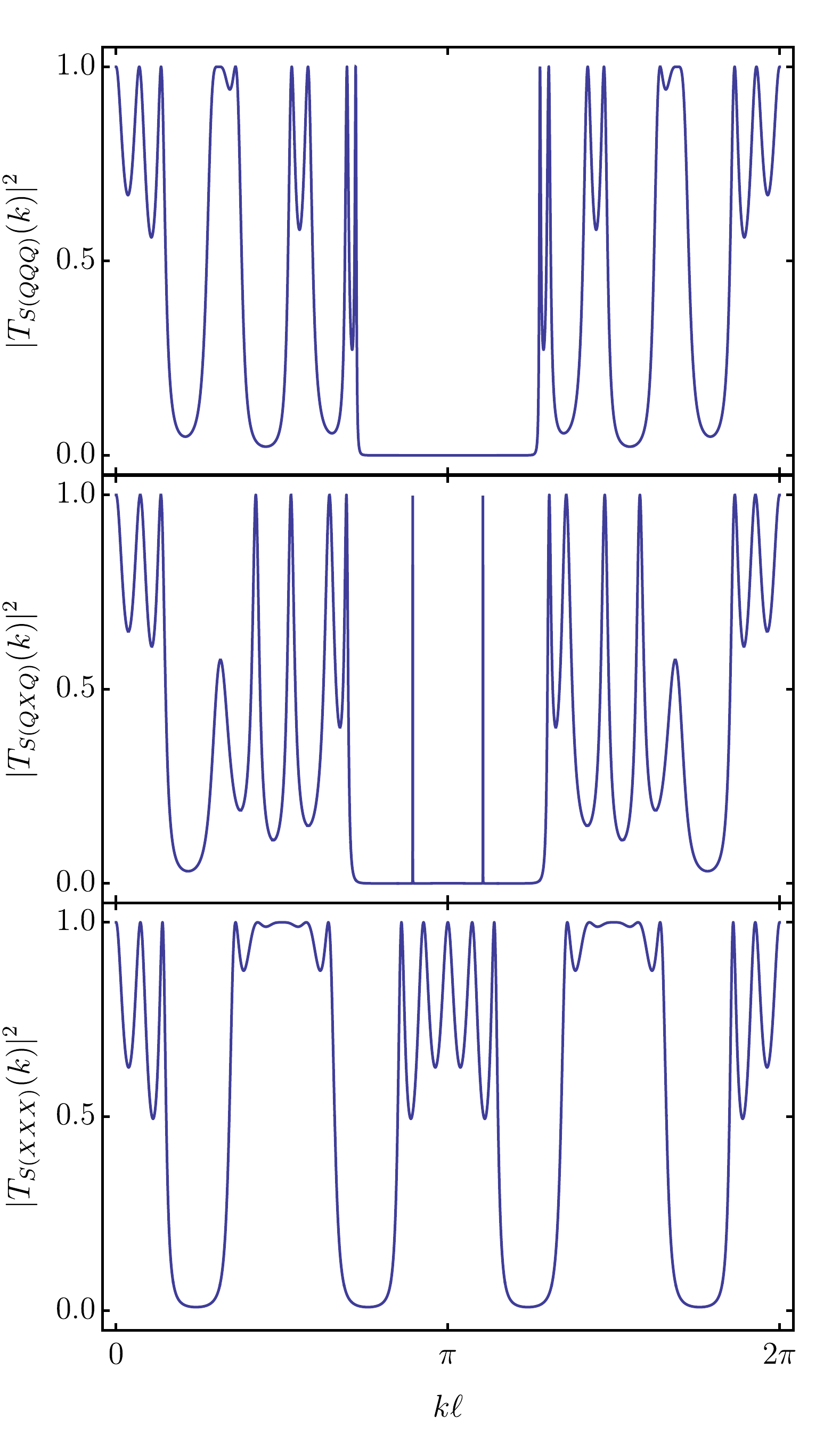}
 \caption[]{(Color online)
   Transmission coefficients of the three compound structures, the
   series arrangements $S(QQQ)$, $S(QXQ)$, and $S(XXX)$, depicted from
   top to bottom, respectively.
 }
  \label{fig:fig8}
\end{figure}

To do this, we add another basic device to the series structure, so we
consider compound structures with three devices.
In this case there are several possibilities and in Fig. \ref{fig:fig8}
we depict the three series arrangements $S(QQQ)$, $S(QXQ)$ and
$S(XXX)$, which are important for the considerations that follow below.
The top and bottom panels in Fig. \ref{fig:fig8} show a behavior which
appeared before, when we compared with Figs. \ref{fig:fig5} and
\ref{fig:fig7}.
In particular, in the top panel in Fig. \ref{fig:fig8}, one
sees that the transmission coefficient vanishes completely in some
interval in $k$ space, so we can also use this in applications of
current interest.
Moreover, the behavior that appears in the middle panel in
Fig. \ref{fig:fig8} reveals an interesting quantum behavior -- the
presence of two very narrow peaks of full transmission inside the
suppression band.
They are very interesting and are consequences of the constructive
quantum interference in the underlying graph, which we further study in
Sec. \ref{sec:r}. 

\subsection{Simple parallel circuits}

Let us now study the case of parallel circuits.
Here the condition that we only have vertices of degree 3
selects some specific combinations of the elementary devices.
The simplest parallel possibilities are the $P(QQ)$, $P(QX)=P(XQ)$ and
$P(XX)$ structures, where $P$ is used to indicate parallel
arrangements.
The $P(QX)$ arrangement, for instance, is constructed as follows:
one puts the $Q$ graph on top of the $X$ graph, without contact, and at
the center of the vertical arrangement, at the left and right one adds
two extra vertices, the one at the left (right) being connected with a
left (right) lead, and then connected with two other edges to keep the
degree 3 condition, one going up to the $Q$ graph, and the other going
down to the $X$ graph.

\begin{figure}[t!]
   \centering
   \includegraphics*[width=\columnwidth]{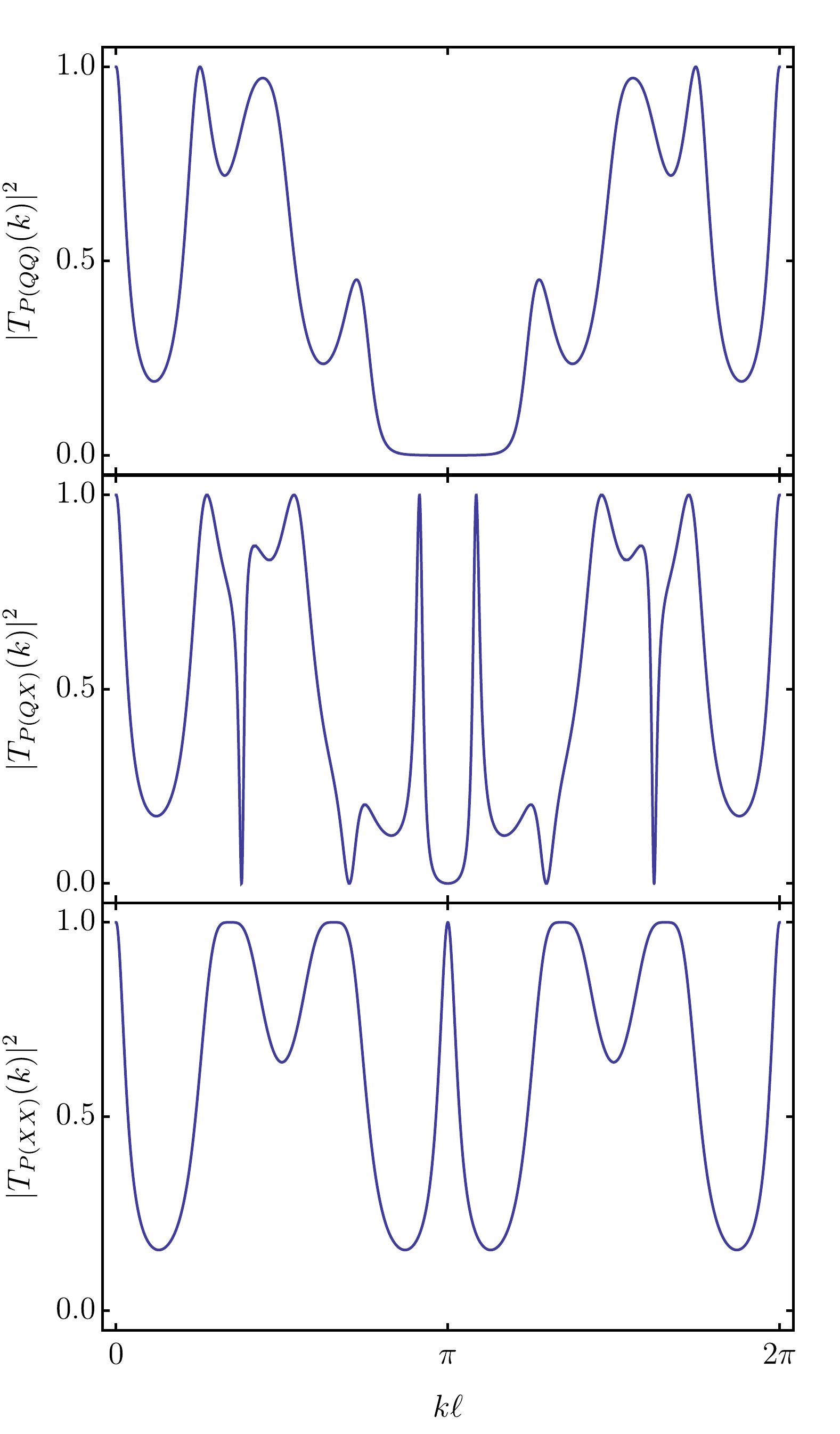}
   \caption[]{(Color online)
     Transmission coefficients  of the three compound structures, the
     parallel arrangements $P(QQ)$, $P(QX)$, and $P(XX)$, depicted from
     top to bottom, respectively.
   }
   \label{fig:fig9}
\end{figure}

We study the three distinct possibilities and in Fig. \ref{fig:fig9} we
depict the transmission coefficients for the three distinct cases.
We note that both the top $P(QQ)$ and bottom $P(XX)$ figures give
results somehow similar to the respective cases in the series
arrangements shown in Fig. \ref{fig:fig7};
compare the top results and the bottom results of both
Figs. \ref{fig:fig7} and \ref{fig:fig9}.
However, the middle panel which describes the $P(QX)$ possibility is
different from the case displayed in the middle panel of
Fig. \ref{fig:fig7}, so we go further and study other compositions.

\subsection{Other arrangements}

The above results suggest that we study other possibilities.
The series and parallel arrangements are more intricate than the
elementary $Q$ and $X$ compositions, and they require more complicated
numerical calculations.
However, if one keeps the condition of vertices of degree 3, there are
several possibilities and we can, for instance, consider the parallel
structures $P(QQ)$,$P(QX)$ and $P(XX)$ in parallel and in series.
Examples are the cases $P(P(QQ)P(QX))$, and $P(P(QX)P(XQ))$, which
represent parallel arrangements of parallel arrangements, etc., and
$S(P(QQ)P(XX)P(QQ))$, which represents a series arrangement of three
parallel arrangements, etc.

\begin{figure}[t!]
   \centering
   \includegraphics*[width=\columnwidth]{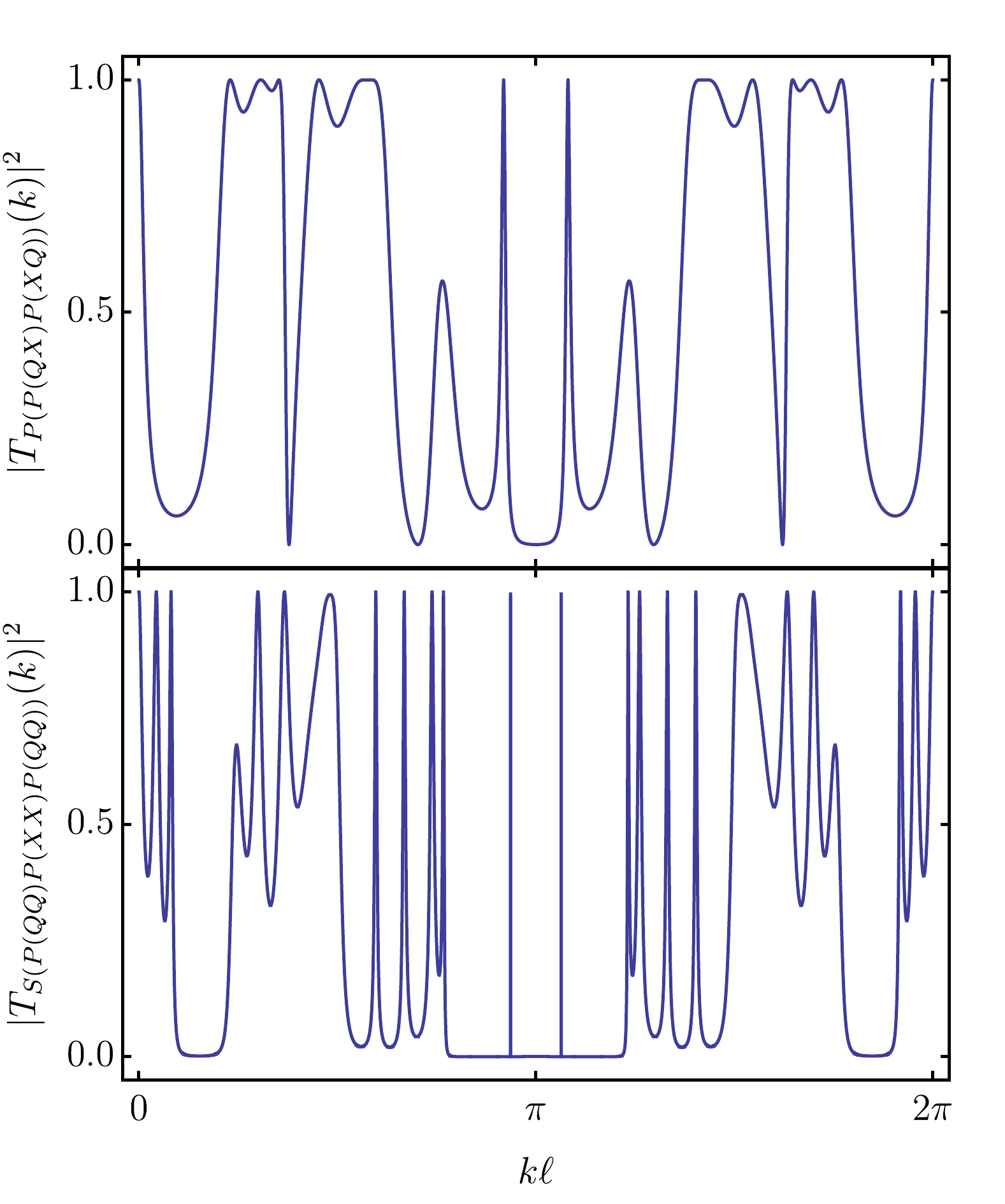}
   \caption[]{(Color online)
     Transmission coefficients of two compound structures, the parallel
     arrangement of two parallel structures,
     $P(P(QX)P(XQ))$, and the series arrangements of three parallel
     structures
     $S(P(QQ)P(XX)P(QQ))$, depicted from top to bottom, respectively.
}
\label{fig:fig10}
\end{figure}

We have studied several cases and, in comparison with the previous
results, we found no qualitatively different behavior.
To exemplify the findings, let us consider for instance the case of a
parallel composition of two parallel compositions and a series
composition with three structures of two parallel compositions.
The results are depicted in Fig. \ref{fig:fig10}, for the cases
$P(P(QX)P(XQ))$ and $S(P(QQ)P(XX)P(QQ))$, respectively.
We note that the transmission coefficients for these new compositions
add no different qualitative effects, in comparison with the previous
results, so we end the calculations of transmission coefficients here.

%%%%%%%%%%%%%%%%%%%%%

\section{Interference}
\label{sec:r}

We see, from the transmission coefficients of the several arrangements
already studied, the appearance of peaks of full transmission in the
region around the center ($k\ell=\pi$) of the periodic region in $k$
space, which we now want to investigate more carefully.

We first focus on the central peak that is displayed with the violet
dotted line in Fig. \ref{fig:fig5}.
We do this by looking at the poles of the Green's function which, for
the $X$ graph depicted in Fig. \ref{fig:fig3}, are all contained in the
roots of the denominator of Eq. \eqref{eq:tX} \cite{PR.647.1.2016}.
Here we identified a pole
at $k\ell =\pi$, so one extends the investigation to the complex plane
to find that this pole has a width that measures $w_{X}=0.54408$.
Similarly, we also confirmed the presence of the pole at $k\ell=\pi$ in
the bottom panel in Fig. \ref{fig:fig7}, but now the width is
$w_{S(XX)}=0.25037$.

The more interesting case appears from the middle panel in
Fig. \ref{fig:fig8} and in the bottom panel in Fig. \ref{fig:fig10}.
There are two similar peaks which engender very
narrow  widths, so we further examine the corresponding Green's function
and find the two poles that appear in the middle panel in
Fig. \ref{fig:fig8}:
they are located at $k\ell=\pi\pm 0.33250$, and have very narrow width,
which obeys $w<0.00030$. They are peaks of full transmission that appear
inside the band of full suppression, and can be interpreted as peaks of constructive quantum interference.
A similar situation appears in the bottom panel in Fig. \ref{fig:fig10}.
The presence of the two peaks symmetrically located around
$k\ell=\pi$ is a consequence of the time reversal symmetry of the
quantum graphs, which does not distinguish the signal entering from the
left (right) and leaving to the right (left).
Note that the time reversal symmetry is present in all graphs here
studied; it shows that $|T(\pi+k\ell)|^2=|T(\pi-k\ell)|^2$, for
$k\ell\in[0,\pi]$, for all the global transmission coefficients.

\section{Discussion}
\label{sec:end}

In this work we studied global transmission properties of some simple
quantum graphs.
We started with diamond and hexagonal graphs and, in the diamond family
of graphs, we compared the transmission coefficients associated with the
$D$ and ${\tilde D}$ arrangements that are depicted in
Figs. \ref{fig:fig1} and \ref{fig:fig3}.
The difference $\Delta_{{\tilde D}D}(k)$ between the two transmissions
appears in Fig. \ref{fig:fig4}, showing that it can be positive or
negative, depending on the wave number of the incoming signal.
For the $D$ and ${\tilde D}$ graphs, however, the degree of the vertices
and the number of edges are not the same, so we moved on to the
hexagonal family of graphs.
In this case, the $Q$ and $X$ arrangements appeared in the study of
simple graphs that engender nontrivial behavior, and they are all
constructed under the condition that the vertices are of degree 3.
This condition is imposed to circumvent the presence of effects due to
vertices of different degrees that could perhaps complicate the
understanding of the results.

With this condition at hand, we ended up with the two quantum graphs of
the hexagonal family which are displayed in Fig. \ref{fig:fig3},
represented by $Q$ and $X$, respectively.
We calculated the corresponding global transmission coefficients
$|T_{Q}(k)|^2$ and $|T_{X}(k)|^2$ and examined some of their
properties.
In particular, we showed the presence of the effect that the
difference between the two global transmissions $\Delta_{QX}(k)$
displayed in Fig. \ref{fig:fig5} can be positive or negative, depending
on the value of the incoming wave number.
Although this is similar to the case studied before, related to the
diamond graphs $D$ and ${\tilde D}$, here the $Q$ and $X$ graphs
contains the very same number of leads, vertices, and edges, with the
vertices with the same degree and edges with the same length.
In this sense, the $Q$ and $X$ graphs are different because of the
distinct connections among their vertices, which change their
topological structures.

We also found a surprising quantum behavior, which concerns the
presence of a suppression band, that is, a large region in wave number,
where the global transmission probability is fully suppressed
by the $Q$ graph.  
Motivated by this, we explored other possibilities, using the two graphs
as elementary devices that could be added together in series and/or
parallel, to form composed structures.
We studied several arrangements, finding results that can certainly
motivate the construction of the apparatus of current interest for
controlling the transmission probability.
Also surprising, we showed how to compose the elementary devices to find
very narrow peaks of constructive quantum interference inside the
suppression band of the $Q$ device.
We investigated the values and widths of these peaks and showed that
they are indeed very narrow.

If one thinks of the two quantum graphs as two elementary devices, it is 
possible to probe them following the lines of
Ref. \cite{PRE.69.056205.2004}, in which experimental and theoretical 
results show that microwave networks can simulate quantum graphs with
time reversal symmetry.
This is an interesting line of investigation, and is further connected 
with another very recent investigation \cite{PRL.122.140503.2019} on 
graphs and possible simulations via microwave networks.
One can also think of considering networks of fibers and splitters, as
considered in \cite{PRL.118.123901.2017}.
In this case, in the simplified version we may say that when a signal
reaches a splitter, it is transmitted towards one of the connected
fibers chosen at random, with the transition probability given in terms
of splitting factors, with the signal flowing as a random
walk on the graph \cite{PRL.118.123901.2017,PRA.99.023841.2019}.

Another important line of research concerns the construction of quantum
devices  at the nanometric scale, simulating the two quantum graphs
$Q$ and $X$ with quantum dots connected by edges and leads; see, e.g.,
Refs. \cite{NM.3.380.2004,NL.8.4523.2008} and references therein.
The idea is to suppose that electrons in the incoming lead reach a
quantum dot from one side and leave the device through the quantum dot
at the outgoing lead on the other side, after interacting with the four
other quantum dots that are arranged to form the two hexagonal graphs
displayed in Fig. \ref{fig:fig3}.
Here the matter flow can be controlled by chemical potentials of
electronic sources that are attached to the left and right leads.
From the practical perspective, the experimental construction of devices
based on quantum dots seems to face another challenging obstacle, which
concerns the graph $X$, that requires two edges that cross without
touching each other. To circumvent this, one has to leave the planar
perspective to build spatial devices.
There is no problem here, if one thinks of modeling microwave
structures like the ones described in \cite{PRE.69.056205.2004,PRL.122.140503.2019}
and also, the fabrication of lattices of optical fibers and splitters in
the form recently suggested in
\cite{PRL.118.123901.2017,PRA.99.023841.2019}.
Another possibility of practical interest is to leave the $Q$ and $X$
arrangements and examine simpler graphs, with the focus on the
construction of simpler quantum devices at the nanometric scale.
The challenge here is to conciliate quantum complexity with geometric
simplicity: complexity that is required for the enhancement of the
quantum interference and simplicity which is welcome for the fabrication
of quantum devices. 

The theoretical perspective engenders other realizations, an interesting
one being the study of more realistic graphs.
Another feasible possibility is the inclusion of potentials along the edges
and/or barriers at the vertices of the quantum graphs.
This can be implemented with the addition of real parameters related to
the potentials added along the edges, as commented on below Eq. \eqref{potential}, and
the strength of the $\delta$-type interaction at the vertices; this last
possibility is controlled by Eqs. \eqref{eq:bc_delta},
\eqref{eq:r_delta} and \eqref{eq:t_delta}, and its realization follows
straightforwardly.
In the case of electronic transport, we can also
add appropriate magnetic fields, which would break the time reversal
symmetry and add new effects.
Another line of investigation concerns the search for other graphs, with
similar properties but distinct topologies, which could suggest the
construction of other experimental devices of direct interest to the
control and manipulation of quantum transmission.

Before ending the work, we mention that, besides the very narrow peaks of
full transmission that we found in this work, we have also observed some
sharp peaks of full suppression in Figs. \ref{fig:fig9} and
\ref{fig:fig10}, and they also pose some issues, in particular related
to their origin and narrowness.
We believe that the peaks of full transmission and suppression deserve
further attention, and an investigation focusing on them is now under
consideration, with special attention to the possibility to relate the
topological structures of the $Q$ and $X$ graphs to the topological
resonances considered in \cite{PRL.110.094101.2013}.

\begin{acknowledgments}
This work was partially supported by the Brazilian agencies Conselho
Nacional de Desenvolvimento Cient\'\i fico e Tecnol\'ogico (CNPq),
Funda\c{c}\~{a}o Arauc\'{a}ria (FAPPR, Grant No. 09/2016),
CNPq INCT-IQ (Grant No. 465469/2014-0), and
Para\'iba State Research Foundation (FAPESQ-PB, Grant No. 0015/2019).
It was also financed in part by the Coordena\c{c}\~{a}o de
Aperfei\c{c}oamento de Pessoal de N\'{i}vel Superior (CAPES, Finance
Code 001).
F.M.A. and D.B. also acknowledge CNPq Grants No. 313274/2017-7 (F.M.A),
No. 434134/2018-0 (F.M.A.), No. 306614/2014-6 (D.B.), and 404913/2018-0
(D.B.).
\end{acknowledgments}

\bibliographystyle{apsrev4-1}
\bibliography{bib.bib}

\begin{thebibliography}{29}%
\makeatletter
\providecommand \@ifxundefined [1]{%
 \@ifx{#1\undefined}
}%
\providecommand \@ifnum [1]{%
 \ifnum #1\expandafter \@firstoftwo
 \else \expandafter \@secondoftwo
 \fi
}%
\providecommand \@ifx [1]{%
 \ifx #1\expandafter \@firstoftwo
 \else \expandafter \@secondoftwo
 \fi
}%
\providecommand \natexlab [1]{#1}%
\providecommand \enquote  [1]{``#1''}%
\providecommand \bibnamefont  [1]{#1}%
\providecommand \bibfnamefont [1]{#1}%
\providecommand \citenamefont [1]{#1}%
\providecommand \href@noop [0]{\@secondoftwo}%
\providecommand \href [0]{\begingroup \@sanitize@url \@href}%
\providecommand \@href[1]{\@@startlink{#1}\@@href}%
\providecommand \@@href[1]{\endgroup#1\@@endlink}%
\providecommand \@sanitize@url [0]{\catcode `\\12\catcode `\$12\catcode
  `\&12\catcode `\#12\catcode `\^12\catcode `\_12\catcode `\%12\relax}%
\providecommand \@@startlink[1]{}%
\providecommand \@@endlink[0]{}%
\providecommand \url  [0]{\begingroup\@sanitize@url \@url }%
\providecommand \@url [1]{\endgroup\@href {#1}{\urlprefix }}%
\providecommand \urlprefix  [0]{URL }%
\providecommand \Eprint [0]{\href }%
\providecommand \doibase [0]{http://dx.doi.org/}%
\providecommand \selectlanguage [0]{\@gobble}%
\providecommand \bibinfo  [0]{\@secondoftwo}%
\providecommand \bibfield  [0]{\@secondoftwo}%
\providecommand \translation [1]{[#1]}%
\providecommand \BibitemOpen [0]{}%
\providecommand \bibitemStop [0]{}%
\providecommand \bibitemNoStop [0]{.\EOS\space}%
\providecommand \EOS [0]{\spacefactor3000\relax}%
\providecommand \BibitemShut  [1]{\csname bibitem#1\endcsname}%
\let\auto@bib@innerbib\@empty
%</preamble>
\bibitem [{\citenamefont {Kottos}\ and\ \citenamefont
  {Smilansky}(1999)}]{AoP.274.76.1999}%
  \BibitemOpen
  \bibfield  {author} {\bibinfo {author} {\bibfnamefont {T.}~\bibnamefont
  {Kottos}}\ and\ \bibinfo {author} {\bibfnamefont {U.}~\bibnamefont
  {Smilansky}},\ }\href {\doibase 10.1006/aphy.1999.5904} {\bibfield  {journal}
  {\bibinfo  {journal} {Ann. Phys. (NY)}\ }\textbf {\bibinfo {volume} {274}},\
  \bibinfo {pages} {76} (\bibinfo {year} {1999})}\BibitemShut {NoStop}%
\bibitem [{\citenamefont {Gnutzmann}\ and\ \citenamefont
  {Smilansky}(2006)}]{AP.55.527.2006}%
  \BibitemOpen
  \bibfield  {author} {\bibinfo {author} {\bibfnamefont {S.}~\bibnamefont
  {Gnutzmann}}\ and\ \bibinfo {author} {\bibfnamefont {U.}~\bibnamefont
  {Smilansky}},\ }\href {\doibase 10.1080/00018730600908042} {\bibfield
  {journal} {\bibinfo  {journal} {Adv. Phys.}\ }\textbf {\bibinfo {volume}
  {55}},\ \bibinfo {pages} {527} (\bibinfo {year} {2006})}\BibitemShut
  {NoStop}%
\bibitem [{\citenamefont {Hul}\ \emph {et~al.}(2004)\citenamefont {Hul},
  \citenamefont {Bauch}, \citenamefont {Pako\'{n}ski}, \citenamefont
  {Savytskyy}, \citenamefont {\.{Z}yczkowski},\ and\ \citenamefont
  {Sirko}}]{PRE.69.056205.2004}%
  \BibitemOpen
  \bibfield  {author} {\bibinfo {author} {\bibfnamefont {O.}~\bibnamefont
  {Hul}}, \bibinfo {author} {\bibfnamefont {S.}~\bibnamefont {Bauch}}, \bibinfo
  {author} {\bibfnamefont {P.}~\bibnamefont {Pako\'{n}ski}}, \bibinfo {author}
  {\bibfnamefont {N.}~\bibnamefont {Savytskyy}}, \bibinfo {author}
  {\bibfnamefont {K.}~\bibnamefont {\.{Z}yczkowski}}, \ and\ \bibinfo {author}
  {\bibfnamefont {L.}~\bibnamefont {Sirko}},\ }\href {\doibase
  10.1103/PhysRevE.69.056205} {\bibfield  {journal} {\bibinfo  {journal} {Phys.
  Rev. E}\ }\textbf {\bibinfo {volume} {69}},\ \bibinfo {pages} {056205}
  (\bibinfo {year} {2004})}\BibitemShut {NoStop}%
\bibitem [{\citenamefont {Dick}\ \emph {et~al.}(2004)\citenamefont {Dick},
  \citenamefont {Deppert}, \citenamefont {Larsson}, \citenamefont
  {M{\aa}rtensson}, \citenamefont {Seifert}, \citenamefont {Wallenberg},\ and\
  \citenamefont {Samuelson}}]{NM.3.380.2004}%
  \BibitemOpen
  \bibfield  {author} {\bibinfo {author} {\bibfnamefont {K.~A.}\ \bibnamefont
  {Dick}}, \bibinfo {author} {\bibfnamefont {K.}~\bibnamefont {Deppert}},
  \bibinfo {author} {\bibfnamefont {M.~W.}\ \bibnamefont {Larsson}}, \bibinfo
  {author} {\bibfnamefont {T.}~\bibnamefont {M{\aa}rtensson}}, \bibinfo
  {author} {\bibfnamefont {W.}~\bibnamefont {Seifert}}, \bibinfo {author}
  {\bibfnamefont {L.~R.}\ \bibnamefont {Wallenberg}}, \ and\ \bibinfo {author}
  {\bibfnamefont {L.}~\bibnamefont {Samuelson}},\ }\href {\doibase
  10.1038/nmat1133} {\bibfield  {journal} {\bibinfo  {journal} {Nat. Mater.}\
  }\textbf {\bibinfo {volume} {3}},\ \bibinfo {pages} {380} (\bibinfo {year}
  {2004})}\BibitemShut {NoStop}%
\bibitem [{\citenamefont {Heo}\ \emph {et~al.}(2008)\citenamefont {Heo},
  \citenamefont {Cho}, \citenamefont {Yang}, \citenamefont {Kim}, \citenamefont
  {Lee}, \citenamefont {Lee}, \citenamefont {Kwon}, \citenamefont {Lee},
  \citenamefont {Jo}, \citenamefont {Choi}, \citenamefont {Hyeon},\ and\
  \citenamefont {Hong}}]{NL.8.4523.2008}%
  \BibitemOpen
  \bibfield  {author} {\bibinfo {author} {\bibfnamefont {K.}~\bibnamefont
  {Heo}}, \bibinfo {author} {\bibfnamefont {E.}~\bibnamefont {Cho}}, \bibinfo
  {author} {\bibfnamefont {J.-E.}\ \bibnamefont {Yang}}, \bibinfo {author}
  {\bibfnamefont {M.-H.}\ \bibnamefont {Kim}}, \bibinfo {author} {\bibfnamefont
  {M.}~\bibnamefont {Lee}}, \bibinfo {author} {\bibfnamefont {B.~Y.}\
  \bibnamefont {Lee}}, \bibinfo {author} {\bibfnamefont {S.~G.}\ \bibnamefont
  {Kwon}}, \bibinfo {author} {\bibfnamefont {M.-S.}\ \bibnamefont {Lee}},
  \bibinfo {author} {\bibfnamefont {M.-H.}\ \bibnamefont {Jo}}, \bibinfo
  {author} {\bibfnamefont {H.-J.}\ \bibnamefont {Choi}}, \bibinfo {author}
  {\bibfnamefont {T.}~\bibnamefont {Hyeon}}, \ and\ \bibinfo {author}
  {\bibfnamefont {S.}~\bibnamefont {Hong}},\ }\href {\doibase
  10.1021/nl802570m} {\bibfield  {journal} {\bibinfo  {journal} {Nano Lett.}\
  }\textbf {\bibinfo {volume} {8}},\ \bibinfo {pages} {4523} (\bibinfo {year}
  {2008})}\BibitemShut {NoStop}%
\bibitem [{\citenamefont {Berkolaiko}\ and\ \citenamefont
  {Kuchment}(2012)}]{Book.2012.Berkolaiko}%
  \BibitemOpen
  \bibfield  {author} {\bibinfo {author} {\bibfnamefont {G.}~\bibnamefont
  {Berkolaiko}}\ and\ \bibinfo {author} {\bibfnamefont {P.}~\bibnamefont
  {Kuchment}},\ }\href@noop {} {\emph {\bibinfo {title} {Introduction to
  Quantum Graphs}}}\ (\bibinfo  {publisher} {American Mathematical Society},\
  \bibinfo {year} {2012})\BibitemShut {NoStop}%
\bibitem [{\citenamefont {Schmidt}\ \emph {et~al.}(2003)\citenamefont
  {Schmidt}, \citenamefont {Cheng},\ and\ \citenamefont
  {da~Luz}}]{JPA.36.545.2003}%
  \BibitemOpen
  \bibfield  {author} {\bibinfo {author} {\bibfnamefont {A.~G.~M.}\
  \bibnamefont {Schmidt}}, \bibinfo {author} {\bibfnamefont {B.~K.}\
  \bibnamefont {Cheng}}, \ and\ \bibinfo {author} {\bibfnamefont {M.~G.~E.}\
  \bibnamefont {da~Luz}},\ }\href {\doibase 10.1088/0305-4470/36/42/L01}
  {\bibfield  {journal} {\bibinfo  {journal} {J. Phys. A}\ }\textbf {\bibinfo
  {volume} {36}},\ \bibinfo {pages} {L545} (\bibinfo {year}
  {2003})}\BibitemShut {NoStop}%
\bibitem [{\citenamefont {Andrade}\ \emph {et~al.}(2016)\citenamefont
  {Andrade}, \citenamefont {Schmidt}, \citenamefont {Vicentini}, \citenamefont
  {Cheng},\ and\ \citenamefont {da~Luz}}]{PR.647.1.2016}%
  \BibitemOpen
  \bibfield  {author} {\bibinfo {author} {\bibfnamefont {F.~M.}\ \bibnamefont
  {Andrade}}, \bibinfo {author} {\bibfnamefont {A.~G.~M.}\ \bibnamefont
  {Schmidt}}, \bibinfo {author} {\bibfnamefont {E.}~\bibnamefont {Vicentini}},
  \bibinfo {author} {\bibfnamefont {B.~K.}\ \bibnamefont {Cheng}}, \ and\
  \bibinfo {author} {\bibfnamefont {M.~G.~E.}\ \bibnamefont {da~Luz}},\ }\href
  {\doibase 10.1016/j.physrep.2016.07.001} {\bibfield  {journal} {\bibinfo
  {journal} {Phys. Rep.}\ }\textbf {\bibinfo {volume} {647}},\ \bibinfo {pages}
  {1} (\bibinfo {year} {2016})}\BibitemShut {NoStop}%
\bibitem [{\citenamefont {Andrade}\ and\ \citenamefont
  {Severini}(2018)}]{PRA.98.062107.2018}%
  \BibitemOpen
  \bibfield  {author} {\bibinfo {author} {\bibfnamefont {F.~M.}\ \bibnamefont
  {Andrade}}\ and\ \bibinfo {author} {\bibfnamefont {S.}~\bibnamefont
  {Severini}},\ }\href {\doibase 10.1103/physreva.98.062107} {\bibfield
  {journal} {\bibinfo  {journal} {Phys. Rev. A}\ }\textbf {\bibinfo {volume}
  {98}},\ \bibinfo {pages} {062107} (\bibinfo {year} {2018})}\BibitemShut
  {NoStop}%
\bibitem [{\citenamefont {Braess}(1968)}]{UOR.12.258.1968}%
  \BibitemOpen
  \bibfield  {author} {\bibinfo {author} {\bibfnamefont {D.}~\bibnamefont
  {Braess}},\ }\href {\doibase 10.1007/bf01918335} {\bibfield  {journal}
  {\bibinfo  {journal} {Unternehmensforschung Operations Research}\ }\textbf
  {\bibinfo {volume} {12}},\ \bibinfo {pages} {258} (\bibinfo {year}
  {1968})}\BibitemShut {NoStop}%
\bibitem [{\citenamefont {Feshbach}(1958)}]{AoP.5.357.1958}%
  \BibitemOpen
  \bibfield  {author} {\bibinfo {author} {\bibfnamefont {H.}~\bibnamefont
  {Feshbach}},\ }\href {\doibase 10.1016/0003-4916(58)90007-1} {\bibfield
  {journal} {\bibinfo  {journal} {Ann. Phys. (NY)}\ }\textbf {\bibinfo {volume}
  {5}},\ \bibinfo {pages} {357} (\bibinfo {year} {1958})}\BibitemShut {NoStop}%
\bibitem [{\citenamefont {Braess}\ \emph {et~al.}(2005)\citenamefont {Braess},
  \citenamefont {Nagurney},\ and\ \citenamefont
  {Wakolbinger}}]{TS.39.446.2005}%
  \BibitemOpen
  \bibfield  {author} {\bibinfo {author} {\bibfnamefont {D.}~\bibnamefont
  {Braess}}, \bibinfo {author} {\bibfnamefont {A.}~\bibnamefont {Nagurney}}, \
  and\ \bibinfo {author} {\bibfnamefont {T.}~\bibnamefont {Wakolbinger}},\
  }\href {\doibase 10.1287/trsc.1050.0127} {\bibfield  {journal} {\bibinfo
  {journal} {Transportation Science}\ }\textbf {\bibinfo {volume} {39}},\
  \bibinfo {pages} {446} (\bibinfo {year} {2005})}\BibitemShut {NoStop}%
\bibitem [{\citenamefont {Youn}\ \emph {et~al.}(2008)\citenamefont {Youn},
  \citenamefont {Gastner},\ and\ \citenamefont {Jeong}}]{PRL.101.128701.2008}%
  \BibitemOpen
  \bibfield  {author} {\bibinfo {author} {\bibfnamefont {H.}~\bibnamefont
  {Youn}}, \bibinfo {author} {\bibfnamefont {M.~T.}\ \bibnamefont {Gastner}}, \
  and\ \bibinfo {author} {\bibfnamefont {H.}~\bibnamefont {Jeong}},\ }\href
  {\doibase 10.1103/physrevlett.101.128701} {\bibfield  {journal} {\bibinfo
  {journal} {Phys. Rev. Lett.}\ }\textbf {\bibinfo {volume} {101}},\ \bibinfo
  {pages} {128701} (\bibinfo {year} {2008})}\BibitemShut {NoStop}%
\bibitem [{\citenamefont {Pala}\ \emph {et~al.}(2012)\citenamefont {Pala},
  \citenamefont {Baltazar}, \citenamefont {Liu}, \citenamefont {Sellier},
  \citenamefont {Hackens}, \citenamefont {Martins}, \citenamefont {Bayot},
  \citenamefont {Wallart}, \citenamefont {Desplanque},\ and\ \citenamefont
  {Huant}}]{PRL.108.076802.2012}%
  \BibitemOpen
  \bibfield  {author} {\bibinfo {author} {\bibfnamefont {M.~G.}\ \bibnamefont
  {Pala}}, \bibinfo {author} {\bibfnamefont {S.}~\bibnamefont {Baltazar}},
  \bibinfo {author} {\bibfnamefont {P.}~\bibnamefont {Liu}}, \bibinfo {author}
  {\bibfnamefont {H.}~\bibnamefont {Sellier}}, \bibinfo {author} {\bibfnamefont
  {B.}~\bibnamefont {Hackens}}, \bibinfo {author} {\bibfnamefont
  {F.}~\bibnamefont {Martins}}, \bibinfo {author} {\bibfnamefont
  {V.}~\bibnamefont {Bayot}}, \bibinfo {author} {\bibfnamefont
  {X.}~\bibnamefont {Wallart}}, \bibinfo {author} {\bibfnamefont
  {L.}~\bibnamefont {Desplanque}}, \ and\ \bibinfo {author} {\bibfnamefont
  {S.}~\bibnamefont {Huant}},\ }\href {\doibase 10.1103/physrevlett.108.076802}
  {\bibfield  {journal} {\bibinfo  {journal} {Phys. Rev. Lett.}\ }\textbf
  {\bibinfo {volume} {108}},\ \bibinfo {pages} {076802} (\bibinfo {year}
  {2012})}\BibitemShut {NoStop}%
\bibitem [{\citenamefont {Sousa}\ \emph {et~al.}(2013)\citenamefont {Sousa},
  \citenamefont {Chaves}, \citenamefont {Farias},\ and\ \citenamefont
  {Peeters}}]{PRB.88.245417.2013}%
  \BibitemOpen
  \bibfield  {author} {\bibinfo {author} {\bibfnamefont {A.~A.}\ \bibnamefont
  {Sousa}}, \bibinfo {author} {\bibfnamefont {A.}~\bibnamefont {Chaves}},
  \bibinfo {author} {\bibfnamefont {G.~A.}\ \bibnamefont {Farias}}, \ and\
  \bibinfo {author} {\bibfnamefont {F.~M.}\ \bibnamefont {Peeters}},\ }\href
  {\doibase 10.1103/physrevb.88.245417} {\bibfield  {journal} {\bibinfo
  {journal} {Phys. Rev. B}\ }\textbf {\bibinfo {volume} {88}},\ \bibinfo
  {pages} {245417} (\bibinfo {year} {2013})}\BibitemShut {NoStop}%
\bibitem [{\citenamefont {Cohen}\ and\ \citenamefont
  {Horowitz}(1991)}]{N.352.699.1991}%
  \BibitemOpen
  \bibfield  {author} {\bibinfo {author} {\bibfnamefont {J.~E.}\ \bibnamefont
  {Cohen}}\ and\ \bibinfo {author} {\bibfnamefont {P.}~\bibnamefont
  {Horowitz}},\ }\href {\doibase 10.1038/352699a0} {\bibfield  {journal}
  {\bibinfo  {journal} {Nature}\ }\textbf {\bibinfo {volume} {352}},\ \bibinfo
  {pages} {699} (\bibinfo {year} {1991})}\BibitemShut {NoStop}%
\bibitem [{\citenamefont {Penchina}\ and\ \citenamefont
  {Penchina}(2003)}]{AJP.71.479.2003}%
  \BibitemOpen
  \bibfield  {author} {\bibinfo {author} {\bibfnamefont {C.~M.}\ \bibnamefont
  {Penchina}}\ and\ \bibinfo {author} {\bibfnamefont {L.~J.}\ \bibnamefont
  {Penchina}},\ }\href {\doibase 10.1119/1.1538553} {\bibfield  {journal}
  {\bibinfo  {journal} {Am. J. Phys.}\ }\textbf {\bibinfo {volume} {71}},\
  \bibinfo {pages} {479} (\bibinfo {year} {2003})}\BibitemShut {NoStop}%
\bibitem [{\citenamefont {Barbosa}\ \emph {et~al.}(2014)\citenamefont
  {Barbosa}, \citenamefont {Bazeia},\ and\ \citenamefont
  {Ramos}}]{PRE.90.042915.2014}%
  \BibitemOpen
  \bibfield  {author} {\bibinfo {author} {\bibfnamefont {A.~L.~R.}\
  \bibnamefont {Barbosa}}, \bibinfo {author} {\bibfnamefont {D.}~\bibnamefont
  {Bazeia}}, \ and\ \bibinfo {author} {\bibfnamefont {J.~G. G.~S.}\
  \bibnamefont {Ramos}},\ }\href {\doibase 10.1103/physreve.90.042915}
  {\bibfield  {journal} {\bibinfo  {journal} {Phys. Rev. E}\ }\textbf {\bibinfo
  {volume} {90}},\ \bibinfo {pages} {042915} (\bibinfo {year}
  {2014})}\BibitemShut {NoStop}%
\bibitem [{\citenamefont {Waltner}\ and\ \citenamefont
  {Smilansky}(2013)}]{APPA.124.1087.2013}%
  \BibitemOpen
  \bibfield  {author} {\bibinfo {author} {\bibfnamefont {D.}~\bibnamefont
  {Waltner}}\ and\ \bibinfo {author} {\bibfnamefont {U.}~\bibnamefont
  {Smilansky}},\ }\href {\doibase 10.12693/APhysPolA.124.1087} {\bibfield
  {journal} {\bibinfo  {journal} {Acta. Phys. Pol. A}\ }\textbf {\bibinfo
  {volume} {124}},\ \bibinfo {pages} {1087} (\bibinfo {year}
  {2013})}\BibitemShut {NoStop}%
\bibitem [{\citenamefont {Harrison}\ \emph {et~al.}(2007)\citenamefont
  {Harrison}, \citenamefont {Smilansky},\ and\ \citenamefont
  {Winn}}]{JPA.40.14181.2007}%
  \BibitemOpen
  \bibfield  {author} {\bibinfo {author} {\bibfnamefont {J.~M.}\ \bibnamefont
  {Harrison}}, \bibinfo {author} {\bibfnamefont {U.}~\bibnamefont {Smilansky}},
  \ and\ \bibinfo {author} {\bibfnamefont {B.}~\bibnamefont {Winn}},\ }\href
  {\doibase 10.1088/1751-8113/40/47/010} {\bibfield  {journal} {\bibinfo
  {journal} {J. Phys. A}\ }\textbf {\bibinfo {volume} {40}},\ \bibinfo {pages}
  {14181} (\bibinfo {year} {2007})}\BibitemShut {NoStop}%
\bibitem [{\citenamefont {Saito}\ \emph {et~al.}(1998)\citenamefont {Saito},
  \citenamefont {Dresselhaus},\ and\ \citenamefont
  {Dresselhaus}}]{Book.1998.Saito}%
  \BibitemOpen
  \bibfield  {author} {\bibinfo {author} {\bibfnamefont {R.}~\bibnamefont
  {Saito}}, \bibinfo {author} {\bibfnamefont {G.}~\bibnamefont {Dresselhaus}},
  \ and\ \bibinfo {author} {\bibfnamefont {M.~S.}\ \bibnamefont
  {Dresselhaus}},\ }\href {\doibase 10.1142/p080} {\emph {\bibinfo {title}
  {Physical Properties of Carbon Nanotubes}}}\ (\bibinfo  {publisher} {Imperial
  College Press},\ \bibinfo {year} {1998})\BibitemShut {NoStop}%
\bibitem [{\citenamefont {Neto}\ \emph {et~al.}(2009)\citenamefont {Neto},
  \citenamefont {Guinea}, \citenamefont {Peres}, \citenamefont {Novoselov},\
  and\ \citenamefont {Geim}}]{RMP.81.109.2009}%
  \BibitemOpen
  \bibfield  {author} {\bibinfo {author} {\bibfnamefont {A.~H.~C.}\
  \bibnamefont {Neto}}, \bibinfo {author} {\bibfnamefont {F.}~\bibnamefont
  {Guinea}}, \bibinfo {author} {\bibfnamefont {N.~M.~R.}\ \bibnamefont
  {Peres}}, \bibinfo {author} {\bibfnamefont {K.~S.}\ \bibnamefont
  {Novoselov}}, \ and\ \bibinfo {author} {\bibfnamefont {A.~K.}\ \bibnamefont
  {Geim}},\ }\href {\doibase 10.1103/revmodphys.81.109} {\bibfield  {journal}
  {\bibinfo  {journal} {Rev. Mod. Phys.}\ }\textbf {\bibinfo {volume} {81}},\
  \bibinfo {pages} {109} (\bibinfo {year} {2009})}\BibitemShut {NoStop}%
\bibitem [{\citenamefont {Gilje}\ \emph {et~al.}(2007)\citenamefont {Gilje},
  \citenamefont {Han}, \citenamefont {Wang}, \citenamefont {Wang},\ and\
  \citenamefont {Kaner}}]{NL.7.3394.2007}%
  \BibitemOpen
  \bibfield  {author} {\bibinfo {author} {\bibfnamefont {S.}~\bibnamefont
  {Gilje}}, \bibinfo {author} {\bibfnamefont {S.}~\bibnamefont {Han}}, \bibinfo
  {author} {\bibfnamefont {M.}~\bibnamefont {Wang}}, \bibinfo {author}
  {\bibfnamefont {K.~L.}\ \bibnamefont {Wang}}, \ and\ \bibinfo {author}
  {\bibfnamefont {R.~B.}\ \bibnamefont {Kaner}},\ }\href {\doibase
  10.1021/nl0717715} {\bibfield  {journal} {\bibinfo  {journal} {Nano Lett.}\
  }\textbf {\bibinfo {volume} {7}},\ \bibinfo {pages} {3394} (\bibinfo {year}
  {2007})}\BibitemShut {NoStop}%
\bibitem [{\citenamefont {Diestel}(2010)}]{Book.2010.Diestel}%
  \BibitemOpen
  \bibfield  {author} {\bibinfo {author} {\bibfnamefont {R.}~\bibnamefont
  {Diestel}},\ }\href {http://books.google.com.mt/books?id=NvRXJSl9hUUC} {\emph
  {\bibinfo {title} {Graph Theory}}},\ \bibinfo {edition} {4th}\ ed.,\ Graduate
  Texts in Mathematics Vol. 173\ (\bibinfo  {publisher} {Springer},\ \bibinfo
  {year} {2010})\BibitemShut {NoStop}%
\bibitem [{\citenamefont {Exner}(1995)}]{PRL.74.3503.1995}%
  \BibitemOpen
  \bibfield  {author} {\bibinfo {author} {\bibfnamefont {P.}~\bibnamefont
  {Exner}},\ }\href {\doibase 10.1103/PhysRevLett.74.3503} {\bibfield
  {journal} {\bibinfo  {journal} {Phys. Rev. Lett.}\ }\textbf {\bibinfo
  {volume} {74}},\ \bibinfo {pages} {3503} (\bibinfo {year}
  {1995})}\BibitemShut {NoStop}%
\bibitem [{\citenamefont {{\L}awniczak}\ \emph {et~al.}(2019)\citenamefont
  {{\L}awniczak}, \citenamefont {Lipovsk{\'{y}}},\ and\ \citenamefont
  {Sirko}}]{PRL.122.140503.2019}%
  \BibitemOpen
  \bibfield  {author} {\bibinfo {author} {\bibfnamefont {M.}~\bibnamefont
  {{\L}awniczak}}, \bibinfo {author} {\bibfnamefont {J.}~\bibnamefont
  {Lipovsk{\'{y}}}}, \ and\ \bibinfo {author} {\bibfnamefont {L.}~\bibnamefont
  {Sirko}},\ }\href {\doibase 10.1103/physrevlett.122.140503} {\bibfield
  {journal} {\bibinfo  {journal} {Phys. Rev. Lett.}\ }\textbf {\bibinfo
  {volume} {122}},\ \bibinfo {pages} {140503} (\bibinfo {year}
  {2019})}\BibitemShut {NoStop}%
\bibitem [{\citenamefont {Lepri}\ \emph {et~al.}(2017)\citenamefont {Lepri},
  \citenamefont {Trono},\ and\ \citenamefont
  {Giacomelli}}]{PRL.118.123901.2017}%
  \BibitemOpen
  \bibfield  {author} {\bibinfo {author} {\bibfnamefont {S.}~\bibnamefont
  {Lepri}}, \bibinfo {author} {\bibfnamefont {C.}~\bibnamefont {Trono}}, \ and\
  \bibinfo {author} {\bibfnamefont {G.}~\bibnamefont {Giacomelli}},\ }\href
  {\doibase 10.1103/physrevlett.118.123901} {\bibfield  {journal} {\bibinfo
  {journal} {Phys. Rev. Lett.}\ }\textbf {\bibinfo {volume} {118}},\ \bibinfo
  {pages} {123901} (\bibinfo {year} {2017})}\BibitemShut {NoStop}%
\bibitem [{\citenamefont {Giacomelli}\ \emph {et~al.}(2019)\citenamefont
  {Giacomelli}, \citenamefont {Lepri},\ and\ \citenamefont
  {Trono}}]{PRA.99.023841.2019}%
  \BibitemOpen
  \bibfield  {author} {\bibinfo {author} {\bibfnamefont {G.}~\bibnamefont
  {Giacomelli}}, \bibinfo {author} {\bibfnamefont {S.}~\bibnamefont {Lepri}}, \
  and\ \bibinfo {author} {\bibfnamefont {C.}~\bibnamefont {Trono}},\ }\href
  {\doibase 10.1103/physreva.99.023841} {\bibfield  {journal} {\bibinfo
  {journal} {Phys. Rev. A}\ }\textbf {\bibinfo {volume} {99}},\ \bibinfo
  {pages} {023841} (\bibinfo {year} {2019})}\BibitemShut {NoStop}%
\bibitem [{\citenamefont {Gnutzmann}\ \emph {et~al.}(2013)\citenamefont
  {Gnutzmann}, \citenamefont {Schanz},\ and\ \citenamefont
  {Smilansky}}]{PRL.110.094101.2013}%
  \BibitemOpen
  \bibfield  {author} {\bibinfo {author} {\bibfnamefont {S.}~\bibnamefont
  {Gnutzmann}}, \bibinfo {author} {\bibfnamefont {H.}~\bibnamefont {Schanz}}, \
  and\ \bibinfo {author} {\bibfnamefont {U.}~\bibnamefont {Smilansky}},\ }\href
  {\doibase 10.1103/PhysRevLett.110.094101} {\bibfield  {journal} {\bibinfo
  {journal} {Phys. Rev. Lett.}\ }\textbf {\bibinfo {volume} {110}},\ \bibinfo
  {pages} {094101} (\bibinfo {year} {2013})}\BibitemShut {NoStop}%
\end{thebibliography}%

\end{document}